\documentclass[final,prd,amsfonts,amsmath,amssymb,nobibnotes,nofootinbib,eqsecnum,showpacs,showkeys]{revtex4}
\usepackage{type1cm,bm,dcolumn}
\usepackage[dvips]{graphicx,color}
\newcommand{\be}{\begin{equation}}
\newcommand{\ee}{\end{equation}}
\newcommand{\bea}{\begin{eqnarray}}
\newcommand{\eea}{\end{eqnarray}}
\newcommand{\ben}{\begin{enumerate}}
\newcommand{\een}{\end{enumerate}}
\newcommand{\bi}{\begin{itemize}}
\newcommand{\ei}{\end{itemize}}

\newcommand{\Ord}[2]{\mathcal O \left(#1\right)^{#2}}

\begin{document}

\title{Localizing modes of massive fermions and a U(1) gauge field in 
the inflating baby-skyrmion branes}

\author{T\'{e}rence Delsate}
\email{terence.delsate(at)umons.ac.be}
\affiliation{CENTRA, Departamento de F\'{\i}sica, 
Instituto Superior T\'ecnico, Universidade T\'ecnica de Lisboa - UTL,
Av.~Rovisco Pais 1, 1049 Lisboa, Portugal.\\
Theoretical and Mathematical Physics Dpt.,Universit\'{e} de Mons, 
UMons, 20, Place du Parc, 7000 Mons - Belgium.}

\author{Nobuyuki Sawado}
\email{sawado(at)ph.noda.tus.ac.jp}
\affiliation{Department of Physics, Faculty of Science and Technology,
Tokyo University of Science, Noda, Chiba 278-8510, Japan}
\date{\today}

\begin{abstract}
We consider the six dimensional brane world model, where the brane is described by a localized solution 
to the baby-Skyrme model extending in the extradimensions. The branes have a cosmological constant 
modeled by inflating four dimensional slices and we further consider a bulk cosmological constant. 
We focus on the topological number three solutions and discuss the localization mechanism of the 
fermions on the above 3-branes. We discuss interpretation of the model in term of quark third generation mass and in terms of the inflation 
history. We argue that the model can describe various epochs of the early universe
by suitably choosing the 
parameters. We further discuss the localization properties of gauge fields on the brane and argue that this is achieved only for specific values of the 
electromagnetic coupling, providing a quantization to the electric charge.
\end{abstract}

\maketitle

\section{Introduction}
Theories with extradimensions have been expected to solve the hierarchy 
problem and cosmological constant problem. Experimentally unobserved 
extradimensions indicate that the standard model particles and forces 
are confined to a three brane~\cite{ArkaniHamed:1998rs,ArkaniHamed:1998nn,Randall:1999ee,Randall:1999vf}. 
The Randall-Sundram (RS) brane model in five space-time 
dimensions~\cite{Randall:1999ee,Randall:1999vf} shows that the exponential 
warp factor in the metric can generate a large hierarchy of scales. 
The brane theories in six dimensions in models of topological object 
show a very distinct feature towards 
the fine-tuning and negative tension brane problems. 
In the context abelian strings~\cite{Cohen:1999ia,Gregory:1999gv,Gherghetta:2000qi,
Giovannini:2001hh,Peter:2003zg} 
were investigated, showing that they can realize localization of gravity for negative
cosmological constant. 
For the magnetic monopoles, similar compactification was achieved for both positive 
and negative cosmological constant~\cite{Roessl:2002rv}. 

There are two main contexts in which solitons appear in field theories: one is
related to 
the strings and the magnetic monopoles in non-abelian gauge theories, and the others
are kinds of non-linear type models like the skyrmions, hopfions~\cite{Skyrme:1961vq,Faddeev:1996zj}. 
The Skyrme model is known to possess soliton solutions called baby-skyrmions 
in two dimensional space~\cite{Piette:1994jt,Piette:1994ug,Kudryavtsev:1996er}. 
The warped compactification of the two dimensional 
extra space by such baby skyrmions was already studied~\cite{Kodama:2008xm}
for negative bulk cosmological constant,
based on the assumption that the cosmological constant 
inside the three branes is tentatively set to be zero. 
Addressing the non-zero cosmological constant inside the branes has been 
considered for case of the strings~\cite{Brihaye:2006pi} and 
the monopoles~\cite{Brihaye:2006cs}. Along these directions, 
we also have studied the baby-skyrmion brane with both positive brane cosmological constant and a bulk cosmological 
constant~\cite{Brihaye:2010nf}. 
In this paper, we employ these solutions as backgrounds for the fermions. 

Study of fermion and gauge fields localization on topological defects have been extensively 
studied with co-dimension one \cite{Kehagias:2000au,Melfo:2006hh,Ringeval:2001cq,Koley:2008dh,
Hosotani:2006qp,Agashe:2007jb,Liu:2009mga,Guo:2011qt}
and two \cite{RandjbarDaemi:2000cr,Libanov:2000uf,Neronov:2001qv,
RandjbarDaemi:2003qd,Parameswaran:2006db,Aguilar:2006sz,Gogberashvili:2007gg,Zhao:2007aw,Guo:2008ia,Guo:2009gb}. 
Many years ago, particle localization on a domain wall in higher dimensional
space time was already addressed \cite{Rubakov:1983bz,Akama:1982jy}. 
The authors suggested the possibility of localized zero-modes of fermions 
on the one dimensional kink background in 4+1 space-time with Yukawa-type coupling.
Later, localization of chiral fermions on RS scenario was discussed in \cite{Kehagias:2000au}. 
Analysis for the massive fermionic modes was done by Ringeval {\it et.al.}, in \cite{Ringeval:2001cq}
and later several studies have followed \cite{Agashe:2007jb,Liu:2009mga,Guo:2011qt}.
For co-dimension two, the localization on higher dimensional generalizations of the RS model
was studied within the coupling of real scalar fields \cite{RandjbarDaemi:2000cr}.
Many studies have followed and most of them are based on the 
Abelian Higgs, or Higgs mediated models with the chiral fermions. 

The problem of mass hierarchy in the Standard Model (SM) fermions~\cite{Froggatt:1978nt} 
has been discussed in many articles based on the brane worlds~
\cite{ArkaniHamed:1999dc,RandjbarDaemi:2000cr,Dvali:2000ha,Libanov:2000uf,Neronov:2001qv,Hung:2001hw,
Aguilar:2006sz,Gogberashvili:2007gg,Hosotani:2006qp,Agashe:2007jb,Guo:2008ia,Guo:2009gb,Frere:2010ah}
in several mechanisms. For example, 
in \cite{Neronov:2001qv} the fermions have quantum numbers of the rotational momenta which are origin of 
the generation of fermions. 
The authors of \cite{Aguilar:2006sz,Gogberashvili:2007gg} deal with this problem with somewhat different 
approach. Conical singularity of the background branes and the orbital angular momentum 
of the fermions around the branes are the key role for the generation.  
In \cite{Libanov:2000uf,Frere:2010ah} 
hierarchy between the fermionic generations are explained 
in terms of multi-winding number solutions of the complex scalar fields. 
They observed three chiral fermionic zero modes on a topological defect with winding 
number three and finite masses appear the mixing of these zero modes. 
Although any discussion of brane construction is absent in their discussion, 
the idea is promising. 
In \cite{Hung:2001hw}, the authors have taken into account more realistic 
standard model charges. 

Among these, we consider the localization of the fermions on the ``inflating '' 
baby skyrmion branes with topological charge three. 
The localized modes of fermions are 
confirmed through the analysis of spectral flow of the one particle state \cite{Kahana:1984be}. 
According to the Index theorem a nonzero topological charge implies the 
zero modes of the Dirac operator \cite{Atiyah:1980jh}. 
The zero crossing modes are found to be the localized fermions on the brane. 
So the generation of the fermions is defined in terms of the topological 
charge of the skyrmions with a special quantum number called grandspin $K_3$. 
There are different profiles of the zero crossing behavior for different $K_3$, 
and it is the origin of the finite mass in our point of view.
Our main concern of this paper is localization of the fermions inside 
the brane with positive cosmological constant, i.e. the inflating branes. 

We further try to make more sense of such models from a phenomenological point of view. 
Indeed a four dimensional cosmological constant is now admitted, and the model under 
investigation has positive curvature. However, the positive curvature is here modeled 
by inflation, which is clearly suitable for early universe bit might be inaccurate for 
present epoch. We argue that suitable choice of parameters might still give a 
reasonable approximation of the underlying physics thanks to the fact that only he Hubble parameter
enters the (background) equations. Further changes affect the details of the four dimensional internal physics without 
spoiling the localization properties. The present model can be suited to describe more general positive curvature four dimensional branes.
The Hubble parameter was quite large value at the inflating period; it dramatically 
reduced and is almost negligible at the present epoch. 
Interestingly, the history of such change of the Hubble parameter makes a new 
spectral flow behavior which may explain the particle creation in our Universe. 

Next, as a straightforward extension of the study, 
we consider localization of gauge field minimally coupled to the braneworld. 
It is certainly natural to take into account of particle on the brane after the fermions. 
As a result, we could treat the quarks (and/or electrons) and also the photon on the brane
which certainly ensures to get closer to the Standard Model contents. 
We discuss the localization properties of the gauge field as well
as its effective four dimensional interpretation. 
Our findings indicate that the four dimensional field is localized only for very specific values of the 
electromagnetic coupling.
 
Note that we do not consider the backreactions of the brane due to the gauge fields nor the fermions. Somehow, we assume that the matter 
content of the brane has a negligible contribution to the brane energy, 
i.e. that the content is low energy and the extradimensional physics is of high energy,
which seems reasonable. 

This paper is organized as follows. In the next section we briefly describe the 
Einstein-Skyrme system in six dimensions. Several types of solutions are shown 
in Sec.II. 
We discuss with the formalism of the fermions coupling to the skyrmion
solutions in Sec.III. The spectral flow analysis is introduced in this section.   
In Sec IV, we consider the effects of the inflation to the spectra 
of the fermions. In Sec. V, we discuss the localization of massless electromagnetic fields in the background of the brane. We summarizes in Sec.VI.

\section{The gravitating baby-Skyrme Model in six dimensions}
\subsection{The model}
Let us first introduce the notations and remind the model we study. 
The bosonic sector is the same as in our previous study \cite{Brihaye:2010nf} so we briefly summarize it.  
The total action for the gravitating baby-Skyrme model is of the form $S := S_{\text{grav}} + S_{\text{baby}}$.
The gravitational part of the total action
\begin{equation}
S_{\text{grav}} = \int d^{6}x\sqrt{-g}\left( \frac{1}{2\chi_{(6)}}R - \Lambda_{(6)} \right), 
\label{eq:gravity-action}
\end{equation}
is the generalized Einstein-Hilbert gravity action, where $\Lambda_{(6)}$ is the bulk cosmological constant, and $\chi_{(6)} = 8\pi G_{(6)} = 8\pi/M_{(6)}^{4}$.
The second term $S_{\text{baby}}$ in the total action stands for the baby-Skyrme action and is given by
\be
S_{\text{baby}} = \int d^{6}x\sqrt{-g}\Bigl[\frac{\kappa_{2}}{2}(\partial_{M}\bm{n})\cdot(\partial^{M}\bm{n})-\frac{\kappa_{4}}{4}(\partial_{M}\bm{n}\times\partial_{N}\bm{n})^{2}- \kappa_{0}V(\bm{n}) \Bigr].
\label{eq:brane-action}
\ee
Here $\bm{n}$ is a scalar triplet subject to the nonlinear constraint $\bm{n}\cdot\bm{n}=1$, and $V(\bm{n})$ is the potential term with no derivatives of $\bm{n}$.
The coefficients $\kappa_{0,2,4}$ in Eq.(\ref{eq:brane-action}) are the coupling constants of the gravitating baby-Skyrme model.

Assuming axial symmetry for the extradimensions, the metric can be written in the following form
\begin{equation}
	ds^{2} = M^{2}(\rho)g_{\mu\nu}^{(4)}dx^{\mu}dx^{\nu} - d\rho^{2} - l^{2}(\rho)d\theta^{2}
	\label{eq:6D-metric}
\end{equation}
where $\rho \in [0, \infty)$ and $\theta \in [ 0, 2\pi]$ are the coordinates associated with the extra dimensions.

We further model a cosmological constant on the brane by considering the following form of the four dimensional 
subspace (described by $g_{\mu\nu}^{(4)}$ in Eq.\eqref{eq:6D-metric})
\begin{equation}
	ds_{(4)}^{2} = g_{\mu\nu}^{(4)}dx^{\mu}dx^{\nu} = dt^{2} - \delta_{ij}e^{2 H_0 t}dx^{i}dx^{j}
	\label{eq:4D-metric}
\end{equation}
where the constant $H_0$ is so called the Hubble parameter.

The ansatz for the scalar triplet $\bm{n}$ is given by the hedgehog Ansatz:
\begin{equation}
	\bm{n} = (\sin f(\rho)\cos(n\,\theta), \sin f(\rho)\sin(n\,\theta), \cos f(\rho)).
	\label{eq:hedgehog-ansatz}
\end{equation}

Let us note that there are some variations {\cite{Eslami:2000tj}} for choosing the potential term $V(\bm{n})$
in the baby-Skyrme model (\ref{eq:brane-action}).
Here we use the so-called \emph{old} baby skyrmions potential, which reads
\begin{equation}
	V(\bm{n}) = 1 - \bm{n}\cdot\bm{n}^{(\infty)} = 1 + \cos f(\rho),
	\label{eq:potential}
\end{equation}
where $\bm{n}^{(\infty)} = \lim_{\rho\to\infty}\bm{n}(\rho,\theta)$ is the vacuum configuration of the baby-Skyrme model.

\subsection{Field equations, boundary conditions and parameters of the model}
The equations of motions are extensively described in \cite{Brihaye:2010nf} so we skip the details. 
Here we only present the definitions of the reduced variables for the sake of the readers' understanding:
%
%
\begin{equation}
	r := \sqrt{\frac{\kappa_{2}}{\kappa_{4}}}\,\rho,\;\;\;
	L(r) := \sqrt{\frac{\kappa_{2}}{\kappa_{4}}}\,l(r),\;\;\;
	\label{eq:dimensionless}
\end{equation}

and dimensionless parameters
\be
\alpha := \chi_{(6)}\kappa_{2},\ \beta := \Lambda_{(6)}\frac{\kappa_{4}}{\kappa_{2}^{2}},
\ \mu :=\kappa_{0}\frac{\kappa_{4}}{\kappa_{2}^{2}}, \gamma = \frac{\kappa_4}{\kappa_2}H_0^2
\label{eq:dimensionless2}
\ee
Note that it is possible to interpret $\gamma$ as a positive cosmological constant in the 
four dimensional subspace of the full model using the four dimensional effective theory following 
the lines of \cite{Brihaye:2006pi}. Indeed in this case, $g^{(4)}_{ab}$ is such 
that $G^{(4)}_{ab} = 3H^2 g^{(4)}_{ab}$, where $G^{(4)}_{ab}$ is the Einstein tensor computed with $g^{(4)}_{ab}$. 
Note that replacing the four dimensional subspace by another Einstein spacetime satisfying 
$G^{(4)}_{ab} = 3H^2 g^{(4)}_{ab}$, such as the Schwarzschild-de Sitter spacetime, leads to the same equations.

Let us remind the boundary conditions too:
\begin{equation}
	f(0) = -(m-1)\pi,\;\;\; f(\infty) = \pi,
	\label{eq:BC-f}
\end{equation}
where $m, n \in \mathbb{Z}$, for the baby Skyrme field and
\be
L(0)=0,\ L'(0)\equiv \frac{dL(r)}{dr}\Bigr|_{r=0}=1,\ M(0)=1,\ M'(0)\equiv \frac{dM(r)}{dr}\Bigr|_{r=0}=0,
\label{eq:BC-gr}
\ee
for the metric fields.
The above boundary conditions are required for regularity and finiteness of the energy.

Remember that considering the hedgehog Ansatz (\ref{eq:hedgehog-ansatz}) with the boundary condition (\ref{eq:BC-f}) leads to a topological charge (or winding number) given by
\be
N = \frac{1}{4\pi}\int \bm{n}\cdot(\partial_{\rho}\bm{n}\times\partial_{\theta}\bm{n}) d\rho\,d\theta= \frac{n}{2}\left[ 1 + (-1)^{1-m} \right] \in \mathbb{Z}.
\label{eq:winding}
\ee

Another useful quantity is the rescaled Ricci scalar which will be used later and is given by
\be
R = \frac{2 L''}{L}+\frac{8 L' M'}{L M}+\frac{8M''}{M}+\frac{12 (M')^2}{M^2}-\frac{12 \gamma }{M^2}.
\ee

\subsection{Asymptotic solutions}
Here we briefly remind the asymptotic solutions available in the gravitating baby Skyrme model.

For the large asymptotics $r\to\infty$, we can set $f(r)=\pi$,i.e., the baby-Skyrme field is a topological vacuum. 
For $\beta = 0$ we have a cigar-type set of solutions given by
\begin{equation}
	M_{c}(r) = \gamma(r - r_{0}),\;\;\;
	L_{c}(r) = L_{0} \equiv \gamma\,C_{L}.
\label{bginf0}
\end{equation}
Periodic solutions are found for $\beta > 0$:
\begin{equation}
M_{p}(r) = \sqrt{\frac{10\gamma^{2}}{\alpha\beta}}\sin\left(\sqrt{\frac{\alpha\beta}{10}}(r - r_{0})\right),
\;\;\;L_{p}(r) = L_{0}\cos\left(\sqrt{\frac{\alpha\beta}{10}}(r - r_{0})\right).
\label{bginf1}
\end{equation}
For $\beta < 0$ we have diverging solutions
\begin{equation}
M_{d}(r) = \sqrt{\frac{10\gamma^{2}}{-\alpha\beta}}\sinh\left(\sqrt{\frac{-\alpha\beta}{10}}(r - r_{0})\right),
\;\;\;L_{d}(r) = L_{0}\cosh\left(\sqrt{\frac{-\alpha\beta}{10}}(r - r_{0})\right).
\label{bginf-1}
\end{equation}

The near origin behaviour of the functions $f,L,M$ subject to the boundary conditions \eqref{eq:BC-f} and \eqref{eq:BC-gr} is given by
\bea
f(r)&=& -(m-1)\pi + f^{(n)}(0) r^n + \Ord{r}{n+1}, 
\;L(r)=r+\lambda_1\frac{r^3}{3!}+\Ord{r}{4},\nonumber\\
M(r)&=1&+(\lambda_1+\gamma)\frac{r^2}{2!}+\Ord{r}{3},\; \label{asym0m}
\lambda_1=\frac{-2\gamma-\alpha\left(\beta+\mu-{\left(-1\right)}^m\mu\right)}{4},
\eea
where $f^{(n)}$ stands for $n^{\mbox{th}}$ derivative of $f$.
Note that higher order corrections are straightforward to compute.

\subsection{The $N=3$ solutions}
According to \eqref{eq:winding}, the case of topological number three 
can be achieved only for $n=3$ and odd values of $m$. 
The simplest case is of course $(m,n)=(1,3)$ but we also consider some solutions for $(m,n)=(3,3)$. 
In the following analysis, we mainly focus on the case $\beta=0$, i.e. the case of no bulk cosmological constant.

We numerically solve the system of ordinary differential equations with the solver Colsys \cite{colsys}.
Here we report some notable features of our solutions when varying the model parameters. The solutions we built share the common qualitative properties as their $n=2$ counterpart, up to some details which does not affect the underlying physics. For example, 
there is a maximum value of $\gamma$ that becomes a critical point where the solutions cease to exist for each $\mu,\ \alpha$. 
The baby skyrmion somehow shrinks to smaller radii as $\gamma$ grows until it disappears.

\begin{figure}
\includegraphics[scale=.3]{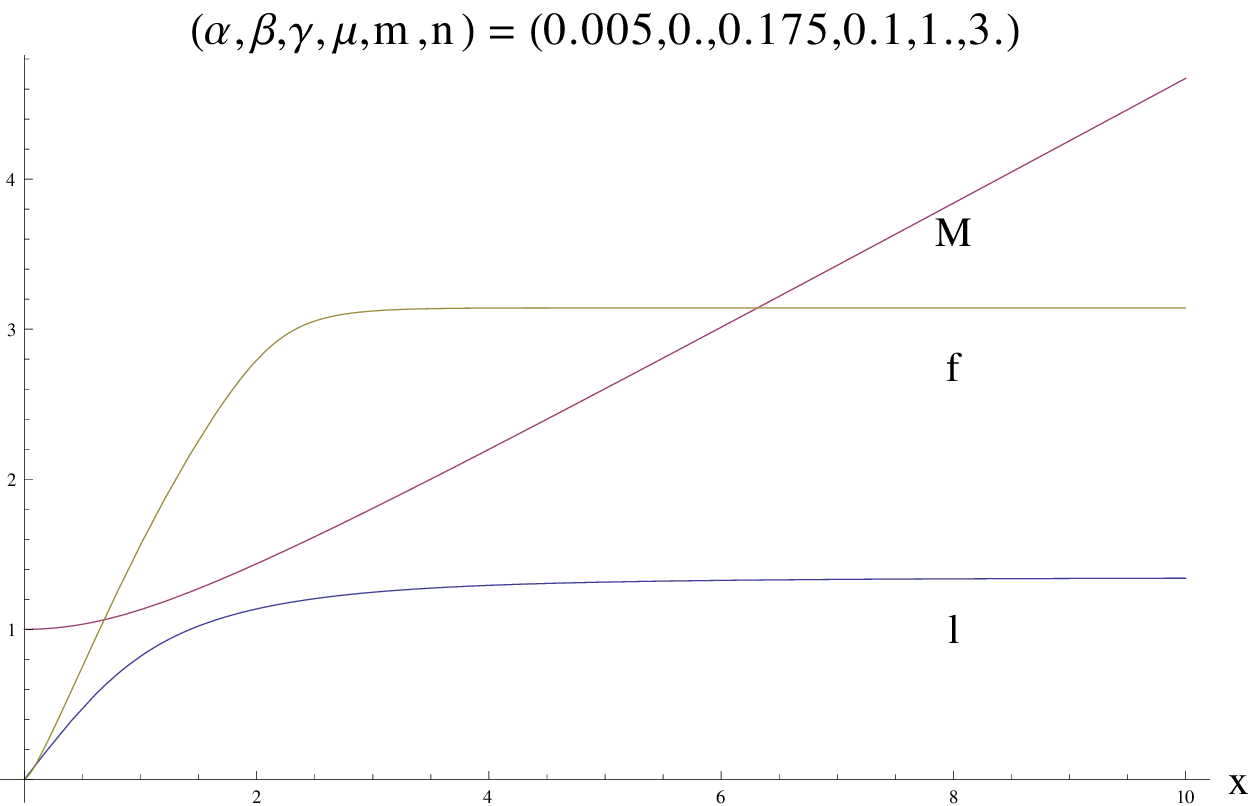}
\caption{A typical solution for $n=3, m=1$.}
\label{fig:sols} 
\end{figure}

\begin{figure}
\includegraphics[scale=.3]{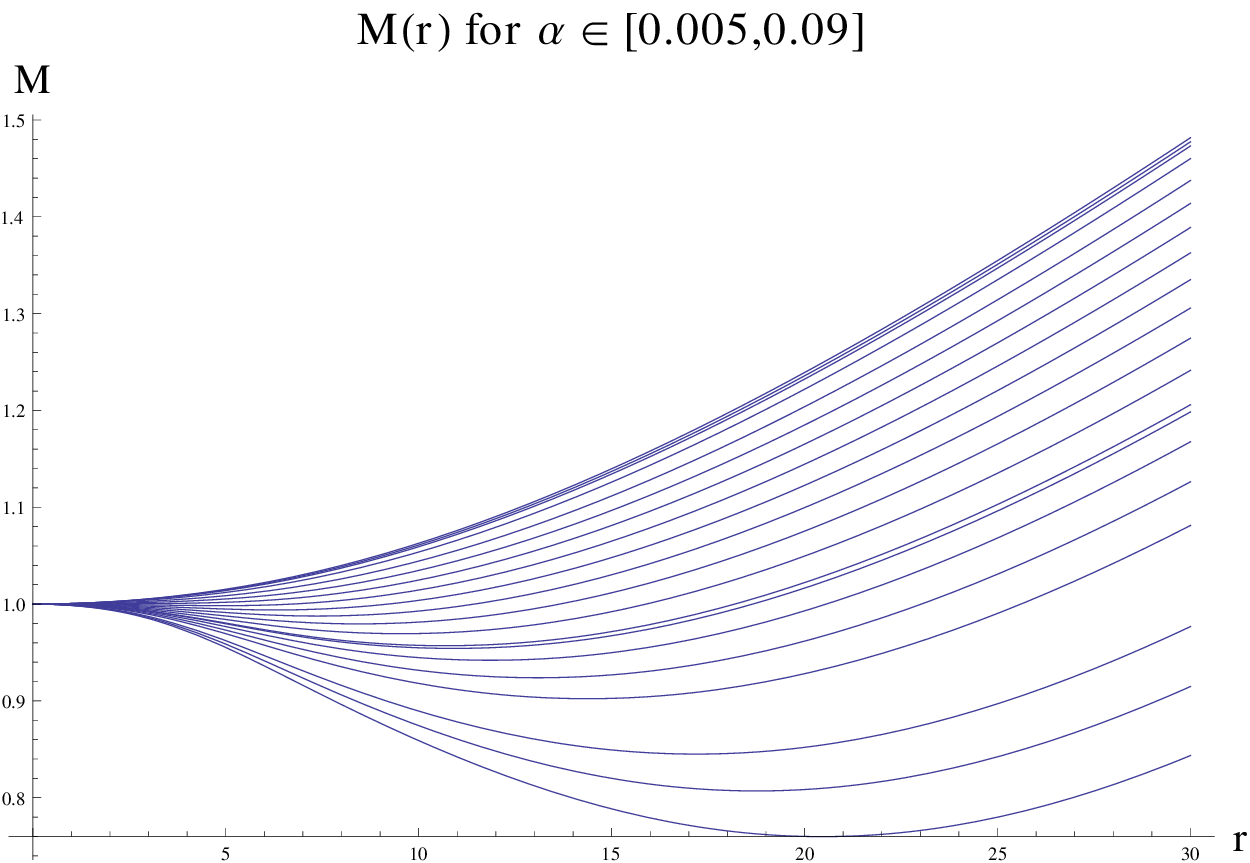}\hspace{0.5cm}\includegraphics[scale=.3]{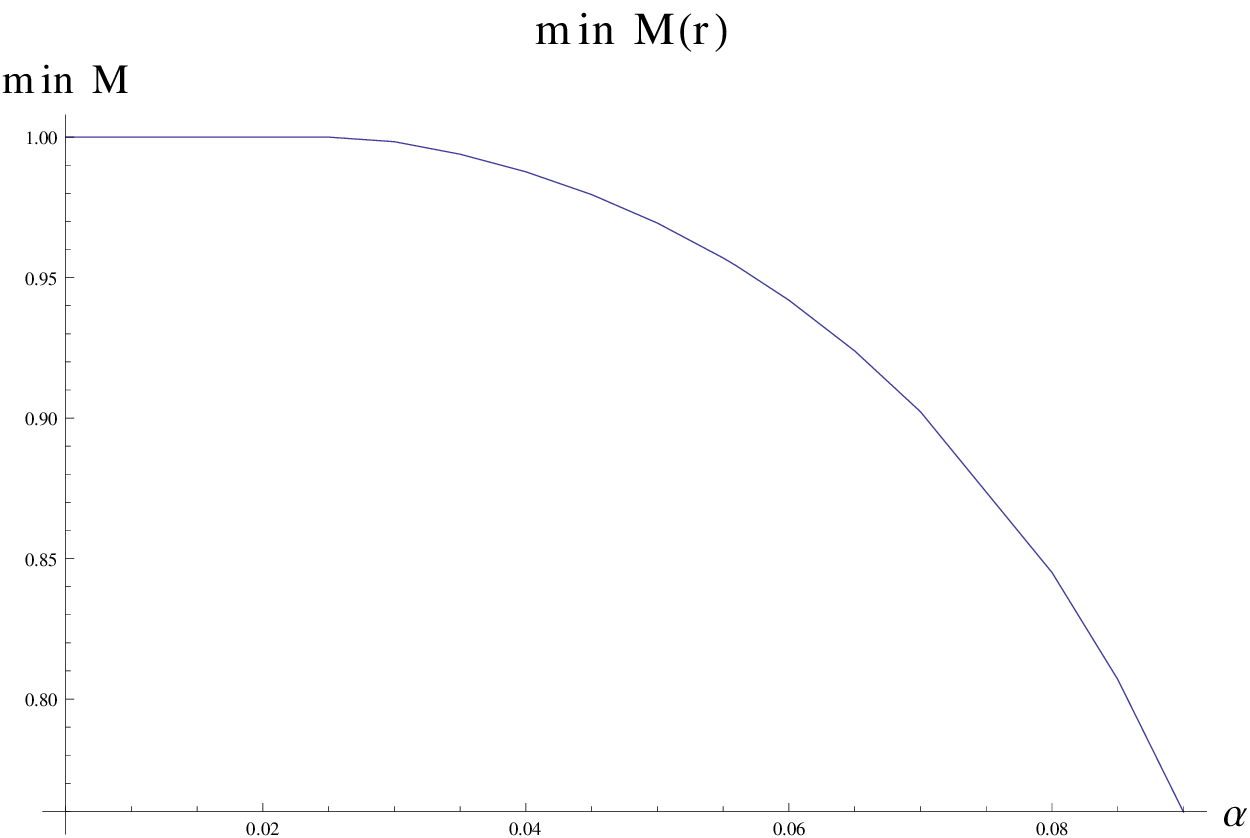}
\caption{For some range of parameters, the warp factor $M$ develops a minimum, it is tempting to assume that the solution ceases to exist when the warp factor 
becomes zero. The parameters of the above solutions are $\gamma=0.001,\ \beta = 0,\ \mu=0.1,\ m=1,\ n=3$ and the range of $\alpha$ is given on the figure.}
\label{fig:minM} 
\end{figure}

We show the solutions with $m=1$ on Fig. \ref{fig:sols}. The metric function $M$ develops a minimum for certain range of parameter $\alpha$ for $(m,n)=(1,3)$. The interpretation is the following: the baby Skyrme wants to shrink the four-dimensional slices up to some 
radii where it exercises its gravitational interaction, but then for larger radius, 
the spacetime is no longer influenced by the brane and the four dimensional slice grows. 
Interestingly, such phenomenon do not occur for $m=3$. This is illustrated on Fig. \ref{fig:minM}.

The matter distribution in the case $(m,n)=(1,3)$ is of a disk shape with more matter at the outside. 
As $\gamma$ increases, even if the root mean square radius decreases, the matter distribution 
becomes dominant on a ring close to the origin but remaining of a disk shape.

\begin{figure}[h]
\includegraphics[scale=.45]{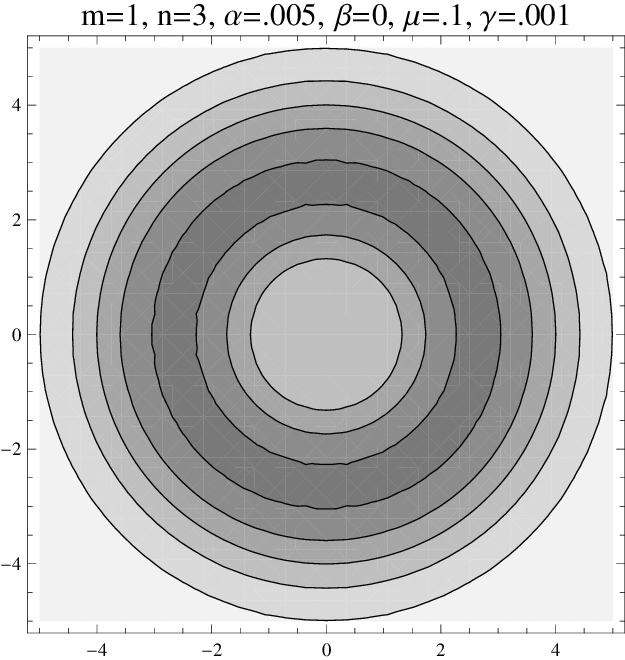}\hspace{1cm}
\includegraphics[scale=.45]{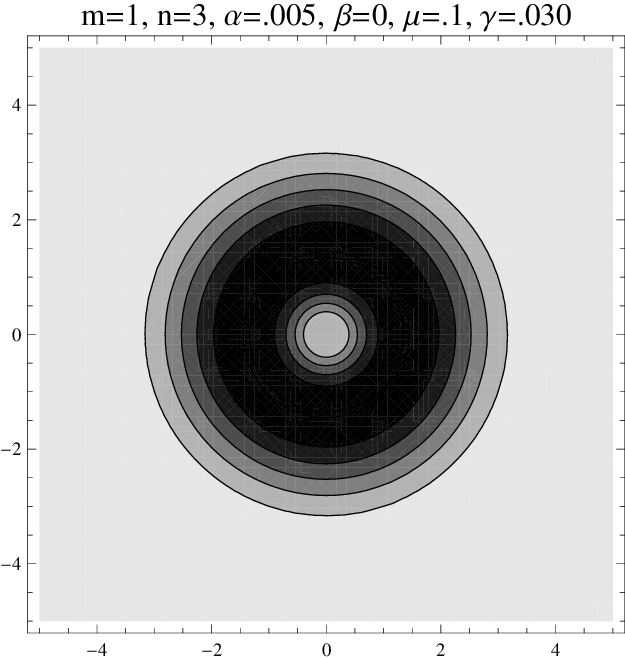}
\caption{Matter distribution for $(m,n)=(1,3)$ in two cases, the center is not vacuum and the ring shape is more pronounced as $\gamma$ increases.}
\label{fig:distribm1n3} 
\end{figure}

\begin{figure}[h]
\includegraphics[scale=.45]{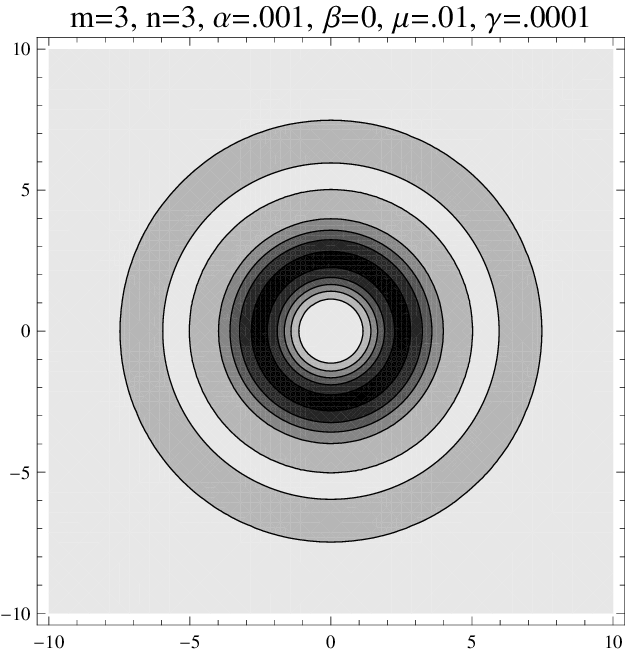}\hspace{.5cm}
\includegraphics[scale=.45]{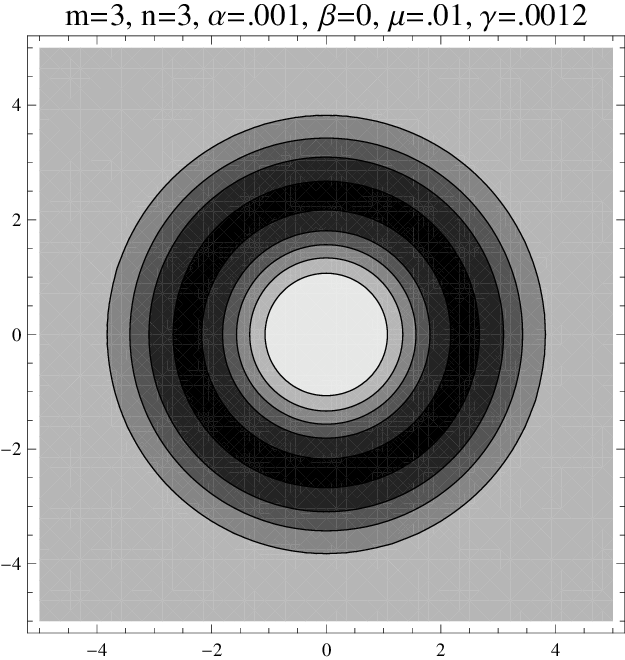}\hspace{.5cm}
\includegraphics[scale=.45]{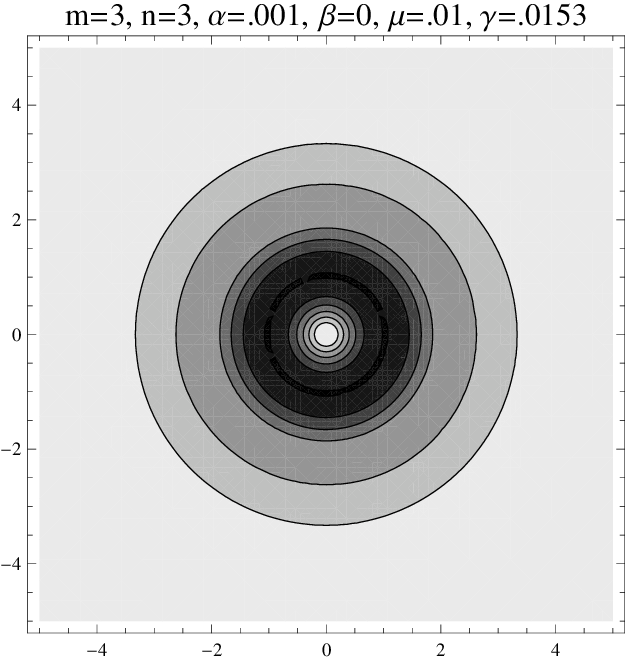}
\caption{Matter distribution for $(m,n)=(3,3)$ in two cases, here the center is vacuum and the distribution is of a ring shape.}
\label{fig:distribm3n3} 
\end{figure}

Instead, for $m=3$, the derivative of the function 
$f$ becomes peaky close to the origin as $\gamma$ increases (again for fixed $\alpha, \ \mu$. 
As a consequence, the matter distribution becomes also peaky on a ring close to the origin. 
Note that in the case $m=3$, the baby skyrmion has indeed a ring shaped matter distribution and no matter is present at the center. 
This is illustrated on Figs. \ref{fig:distribm1n3} and \ref{fig:distribm3n3}.

\section{Massive fermions}
The fermions coupled with the baby-Skyrme field have an isospin doublet structure. Furthermore, we tentatively assume 
that the fermions within an iso-doublet are degenerate. Of course, the assumption is not valid especially for the heavier 
quark sectors. In order to recover it, we would have to introduce additional terms which explicitly break the symmetry. 
Or, different choice of the parameters of the baby-skyrmions for each flavor can exert similar effect, which we shall apply. 
Also, we ignore inter-generation mixing of SM particles.
\subsection{Basic formalism}
We couple the fermions to the baby Skyrme background by virtue of the following action
\begin{eqnarray}
S_{\rm fermion}=\int d^6x \sqrt{-g}[\bar{\Psi}i\Gamma^A D^L_AP_L\Psi+\bar{\Psi}i\Gamma^A D_AP_R\Psi-m\bar{\Psi}(\Phi P_R+\Phi^\dagger P_L)\Psi].
\label{action_S}
\end{eqnarray}
The 6-dimensional gamma matrices $\Gamma^A$ are defined in terms of the {\it vielbein} $e^A_{\hat{a}}$ and of the flat-space $\gamma^{\hat{a}}$ according to $\Gamma^A:=e^A_{\hat{a}}\gamma^{\hat{a}}$.
The covariant derivatives are defined as 
\begin{eqnarray}
D^L_A:=\frac{1}{2}\overleftrightarrow{\partial}_A+\frac{1}{2}\omega_A^{\hat{a}\hat{b}}\sigma_{\hat{a}\hat{b}}-igW_A,~~
D_A:=\frac{1}{2}\overleftrightarrow{\partial}_A+\frac{1}{2}\omega_A^{\hat{a}\hat{b}}\sigma_{\hat{a}\hat{b}},
\end{eqnarray}
where $\omega_A^{\hat{a}\hat{b}}:=\frac{1}{2}e^{\hat{a}B}\nabla_Ae^{\hat{b}}_B$ are the spin connection 
with generators $\sigma_{\hat{a}\hat{b}}:=\frac{1}{4}[\gamma_{\hat{a}},\gamma_{\hat{b}}]$.
The symbol $\overleftrightarrow{\partial}$ acts on spinors according to $\psi\overleftrightarrow{\partial}\phi\equiv \psi\partial \phi-(\partial \psi) \phi$.
Capital latin letters run from $0$ to $5$ and denote the six dimensional space-time index while the hatted small Latin letters range 
from $0$ to $5$ and  corresponds to the flat tangent six dimensional Minkowski space indices. 

We employ the left/right symmetric coupling scheme $\Phi:=\bm{\tau}\cdot\bm{n}, W_A=0$; this form was extensively studied in \cite{Jaroszewicz:1984xw,Carena:1990vy} 
and was applied to the six dimensional brane physics with warped geometry \cite{Kodama:2008xm}.

The vielbein is defined through $g_{AB}=e^{\hat{a}}_Ae_{\hat{a}B}=\eta_{\hat{a}\hat{b}}e^{\hat{a}}_Ae^{\hat{b}}_B$.
We use the following form
\begin{eqnarray}
&&e^{\hat{0}}_0=M(\rho),~~ 
e^{\hat{a}}_k=M(\rho)e^{H_0 t} \delta^{\hat{a}}_k,~~k=1,\cdots,3,\nonumber \\
&&e^{\hat{4}}_\rho=\cos \theta,~~e^{\hat{5}}_\rho=\sin \theta,~~\nonumber \\
&&e^{\hat{4}}_\theta=-l(\rho)\sin\theta,~~e^{\hat{5}}_\theta=l(\rho)\cos\theta\,.
\end{eqnarray}
A straightforward calculation shows that the nonvanishing components of the 
corresponding spin connections are
\begin{eqnarray}
&&\omega^{\hat{0}\hat{4}}_0=-\cos\theta M',~~
\omega^{\hat{a}\hat{4}}_k=-\cos\theta M'e^{H_0 t}\delta^{\hat{a}}_k, 
~~k=1,\cdots,3 \nonumber \\
&&\omega^{\hat{0}\hat{5}}_0=-\sin\theta M',~~
\omega^{\hat{a}\hat{5}}_k=-\sin\theta M'e^{H_0 t}\delta^{\hat{a}}_k, 
~~k=1,\cdots,3 \nonumber \\
&&
\omega^{\hat{4}\hat{5}}_\theta=1-l'.
\end{eqnarray}
The standard (Dirac-Pauli) representation of the gamma matrices in six dimension is given by
\begin{eqnarray}
&&\gamma^{\hat{\mu}}:=
\left(
\begin{array}{cc}
\bar{\gamma}^{\hat{\mu}}& 0 \\
0 & -\bar{\gamma}^{\hat{\mu}} \\
\end{array}
\right),~~
\gamma^{\hat{4}}:=
\left(
\begin{array}{cc}
0 & -iI_4 \\
-iI_4 & 0 \\
\end{array}
\right),~~
\gamma^{\hat{5}}:=
\left(
\begin{array}{cc}
0 & -I_4 \\
I_4 & 0 \\
\end{array}
\right),
\end{eqnarray}
where $\bar{\gamma}^{\hat{\mu}}$ are the standard representation of the usual gamma matrices.
The six dimensional spinor $\Psi$ can be decomposed into a four dimensional and an extradimensional components as
$\Psi(x^\mu,\rho,\theta)=\psi(x^\mu) \otimes (U_1(\rho,\theta),U_2(\rho,\theta))^T$, where $\psi (x^\mu)$, $U_i(\rho,\theta )$ 
are four, two components spinor, respectively. 
In terms of the decomposition, the equations for the extradimensional components are 
\begin{eqnarray}
&&-Me^{-i\theta}\Bigl(\partial_\rho-\frac{i}{l}\partial_\theta-2\frac{\partial_\rho M}{M}
-\frac{\partial_\rho l-1}{2l}\Bigr)U_2+Mm\bm{\tau}\cdot\bm{n}U_1=wU_1 \nonumber \\
&&~~~~Me^{i\theta}\Bigl(\partial_\rho+\frac{i}{l}\partial_\theta-2\frac{\partial_\rho M}{M}
-\frac{\partial_\rho l-1}{2l}\Bigr)U_1-Mm\bm{\tau}\cdot\bm{n}U_2=wU_2
\end{eqnarray}
where $\psi(x^\mu)$ is the solution of the $4$ dimensional Dirac equation
\begin{eqnarray}
(ig^{{\hat{\mu}}{\hat{\nu}}}\bar{\gamma}_{\hat{\mu}}\partial_{\hat{\nu}}+\frac{3}{2}H_0\gamma_{\hat{0}}-w)\psi=0.
\label{diraceq4}
\end{eqnarray}
Therefore, the Dirac equation in six dimensions reduces to a two dimensional eigenproblem where the eigenvalue $w$ is the masses of the fermions 
measured on the four dimensional brane.
The Hubble constant works as a time component of the vector potential $v_0:=\frac{3}{2}V_0$ which simply shifts  
the energy as $E\to E+v_0$.
In terms of the change of spinor $U\equiv (U_1,U_2)^T$ into $u\equiv (u_1,u_2)^T$,
\begin{eqnarray}
u(r,\theta):= 
\exp\Bigl(2\ln M(r)+\frac{1}{2}\ln L(r)-\frac{1}{2}\int^r\frac{dr'}{L(r')}\Bigr)U(r,\theta),
\label{impspinor}
\end{eqnarray}
the eigenproblem then becomes
\begin{eqnarray}
\mathcal H u=\tilde{w}u,
\label{eigenproblem8}
\end{eqnarray}
where the hamiltonian is given by
\begin{eqnarray}
&&\hspace{-0.5cm} \mathcal  H:=
M\left(
\begin{array}{cc}
\tilde{m}\bm{\tau}\cdot\bm{n} &
-e^{-i\theta}\Bigl(\partial_r-i\dfrac{1}{L}\partial_\theta\Bigr)
\\
e^{i\theta}\Bigl(\partial_r+i\dfrac{1}{L}\partial_\theta\Bigr)
&
-\tilde{m}\bm{\tau}\cdot\bm{n} 
\end{array}
\right)\,.
\label{hamiltonian}
\end{eqnarray}
Note that we have introduced dimensionless coupling constant and eigenvalue 
\begin{eqnarray}
\tilde{m}:=\sqrt{\frac{\kappa_4}{\kappa_2}}m,~~~~\tilde{w}:=\sqrt{\frac{\kappa_4}{\kappa_2}}w
\label{dimensionless_mw}
\end{eqnarray}
and also  $L:=\sqrt{\kappa_2/\kappa_4}l$, see (\ref{eq:dimensionless}).

The hamiltonian (\ref{hamiltonian}) is invariant under 
time-reversal transformation defined by ${\cal T}:=i\gamma^{\hat{5}}\otimes\tau^2 C$,  
where $C$ is the charge conjugation operator. 
One can also easily confirm that the hamiltonian $H$ commutes with ``grandspin'' operator given by
\begin{eqnarray}
K_3:=l_3+\frac{\gamma^{\hat{6}}}{2}+n\frac{\tau^3}{2},
\end{eqnarray}
where $l_3:= -i\frac{\partial}{\partial\theta}$ is the orbital angular momentum in the extra space and  
where we introduced $\gamma^{\hat{6}}:=I\otimes \sigma^3$ for convenience~\cite{Kodama:2008xm}.
As a consequence the eigenstates are specified by the magnitude of the grandspin, $i.e.$, 
\begin{eqnarray}
&&K_3=0,\pm 1,\pm 2,\pm 3\cdots,~~{\rm for~odd~} n \nonumber \\
&&K_3=\pm \frac{1}{2},\pm \frac{3}{2},\pm\frac{5}{2},\cdots,~~{\rm for~even~} n\,.
\end{eqnarray}
and it follows from the time-reversal symmetry that one finds that the states of $\pm K_3$ are degenerate in energy. 

\subsection{Asymptotics}
The general form of solutions to (\ref{eigenproblem8}) is given by
\begin{eqnarray}
u=
\left( 
\begin{array}{c}
g_1(r) e^{i(K_3-\frac{1}{2}-\frac{n}{2})\theta} \\
g_2(r)e^{i(K_3-\frac{1}{2}+\frac{n}{2})\theta} \\
h_1(r)e^{i(K_3+\frac{1}{2}-\frac{n}{2})\theta} \\
h_2(r)e^{i(K_3+\frac{1}{2}+\frac{n}{2})\theta} 
\end{array}
\right).
\end{eqnarray}

Far from the origin, the functions follow an exponential behavior that depends on $\beta$; for $\beta=0$, 
\bea
&&g_1\sim Ae^{-\zeta r},h_1\sim e^{-\zeta r},\nonumber\\
&&g_2\sim Be^{-\eta r},h_2\sim e^{-\eta r},
\eea
where
\bea
&&\zeta=\frac{1}{2L_0}+\frac{1}{L_0}\sqrt{\Bigl(K_3-\frac{n}{2}\Bigr)^2+\tilde{m}^2L_0^2},\nonumber \\
&&A=\Bigl[\frac{1}{2\tilde{m}L_0}+\frac{1}{\tilde{m}L_0}\sqrt{\Bigl(K_3-\frac{n}{2}\Bigr)^2+\tilde{m}^2L_0^2}
-\frac{1}{\tilde{m}L_0}\Bigl(K_3+\frac{1}{2}-\frac{n}{2}\Bigr)\Bigr],\nonumber\\
&&\eta=\frac{1}{2L_0}+\frac{1}{L_0}\sqrt{\Bigl(K_3+\frac{n}{2}\Bigr)^2+\tilde{m}^2L_0^2},\nonumber \\
&&B=\Bigl[\frac{1}{2\tilde{m}L_0}+\frac{1}{\tilde{m}L_0}\sqrt{\Bigl(K_3+\frac{n}{2}\Bigr)^2+\tilde{m}^2L_0^2}
-\frac{1}{\tilde{m}L_0}\Bigl(K_3+\frac{1}{2}+\frac{n}{2}\Bigr)\Bigr].
\eea
For $\beta<0$, the solutions are simply $g_i,h_i\sim e^{-\tilde{m}r}$.

Close to the vicinity of the origin, the regular solutions to the linearized equations are 
\begin{eqnarray}
g_1\sim r^{|K_3-\frac{1}{2}-\frac{n}{2}|},~~
g_2\sim r^{|K_3-\frac{1}{2}+\frac{n}{2}|},~~
h_1\sim r^{|K_3+\frac{1}{2}-\frac{n}{2}|},~~
h_2\sim r^{|K_3+\frac{1}{2}+\frac{n}{2}|}
\end{eqnarray}
Therefore, normalized solutions of the eigenequations (\ref{eigenproblem8}) interpolating between the near origin and far region should exist.

\subsection{Schr\"odinger type equation}
The eigeneqaution (\ref{eigenproblem8}) can be recast into a set of Schr\"odinger-like second order differential equation. 
If we eliminate the components $(h_1,h_2)$ , after a lengthy calculation we finally get
\begin{eqnarray}
\left(
\begin{array}{cc}
-\partial_r^2-(\frac{4M'}{M}+\frac{L'}{L})\partial_r+\frac{l^2}{L^2}+V_{11} & W_1\partial_r+V_{12} 
\\
W_2\partial_r+V_{21}
&
-\partial_r^2-(\frac{4M'}{M}+\frac{L'}{L})\partial_r+\frac{(l+n)^2}{L^2}+V_{22}
\end{array}
\right)\,
\left(
\begin{array}{c}
u_1 \\
u_2 
\end{array}
\right)=0
\label{2diffeq-a}
\end{eqnarray}
where $l:=K_3-\frac{1}{2}-\frac{n}{2}$. Here we used the following replacement
\begin{eqnarray}
\left(
\begin{array}{c}
g_1 \\
g_2 
\end{array}
\right)~\to~
\left(
\begin{array}{c}
e^{-\frac{1}{2}\int^r (A_{11}+\frac{1}{L}-R)dr'}u_1 \\
e^{-\frac{1}{2}\int^r (A_{22}+\frac{1}{L}-R)dr'}u_2
\end{array}
\right).
\end{eqnarray}
Similarly for the lower components $h_1,h_2$, we get 
\begin{eqnarray}
\left(
\begin{array}{cc}
-\partial_r^2-(\frac{4M'}{M}+\frac{L'}{L})\partial_r+\frac{(l+1)^2}{L^2}+U_{11} & -W_2\partial_r+U_{12} 
\\
-W_1\partial_r+U_{21}
&
-\partial_r^2-(\frac{4M'}{M}+\frac{L'}{L})\partial_r+\frac{(l+n+1)^2}{L^2}+U_{22}
\end{array}
\right)\,
\left(
\begin{array}{c}
v_1 \\
v_2 
\end{array}
\right)=0\,.
\label{2diffeq-b}
\end{eqnarray}
The explicit forms of $A_{ij},U_{ij},V_{ij},W_i;i,j=1,2$ are summerized in Appendix A.

\subsection{Spectral flow}
Instead of solving the second order differential equations (\ref{2diffeq-a}),(\ref{2diffeq-b}), we
directly find solutions of the original form of the Dirac equation (\ref{eigenproblem8}).
The computational method is based on a plane wave expansion of the spinor and the matrix diagonalization scheme. 
(The detail was described in Sec.III of our previous paper \cite{Kodama:2008xm}.)
 
Interesting property of such isolated bound states is that they dives from positive energy to negative 
if background fields change. This is called the spectral flow or the level crossing picture \cite{Kahana:1984be}.    
The spectral flow is defined as the number of eigenvalues of Dirac Hamiltonian that cross zero from below
minus the number of eigenvalues that cross zero from above when varying the properties of the 
background fields. According to the index theorem, a nonzero topological charge implies zero
modes of the Dirac operator \cite{Atiyah:1980jh}.
The number of flow coincides with the topological charge and zero modes emerge when the flow crosses zero. 
The level crossing picture was extensively studied in the Dirac equation with non-linear chiral background~\cite{Kahana:1984be,Sawado:2004pm}, 
with Higgs field in the Abelian-Higgs model~\cite{Bezrukov:2005rw,Burnier:2006za} and with non-trivial gauge fields (e.g.,instanton,meron)~\cite{Christ:1979zm,Nielsen:1983rb,Kiskis:1978tb}.
The mechanism can be interpreted as a quantum mechanical description of fermion creation/annihilation. 

\begin{figure*}
\includegraphics[height=6.0cm, width=8cm]{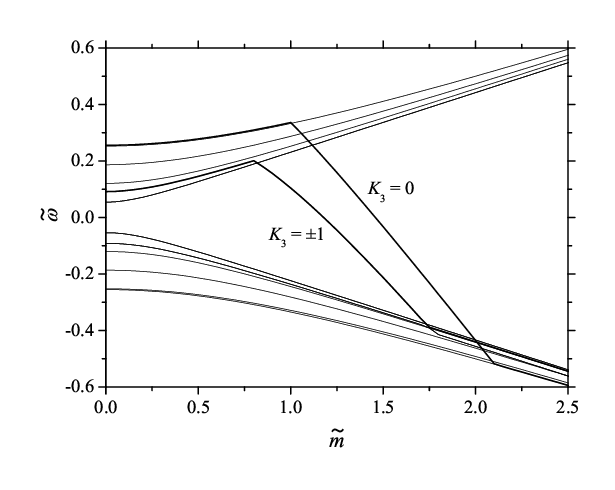}
\includegraphics[height=6.0cm, width=8cm]{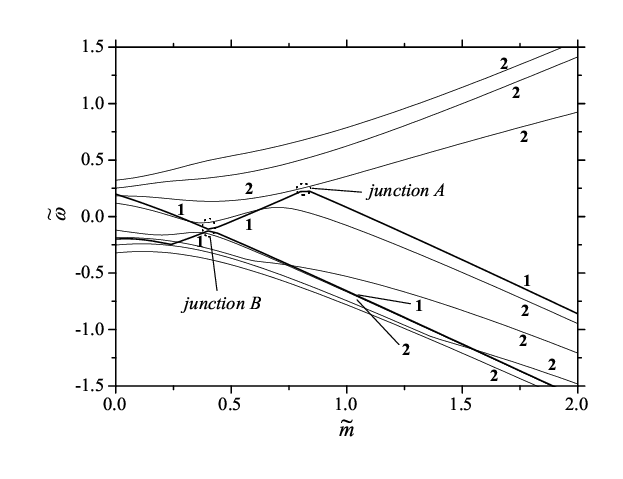}
\caption{\label{fig:spectralflow}The spectral flow with $(m,n)=(1,3)$,$\alpha=0.1,\gamma=0.002$ and $\beta=0.0$ ({\it left}),
and $(m,n)=(3,3)$, $\alpha=0.001,\beta=0.0,\gamma=0.00125$ ({\it right}).
For $(m,n)=(3,3)$, the bold letters beside the lines indicates number of the degeneracy.  
}
\end{figure*}

\begin{figure*}
\includegraphics[height=5.0cm, width=7.5cm]{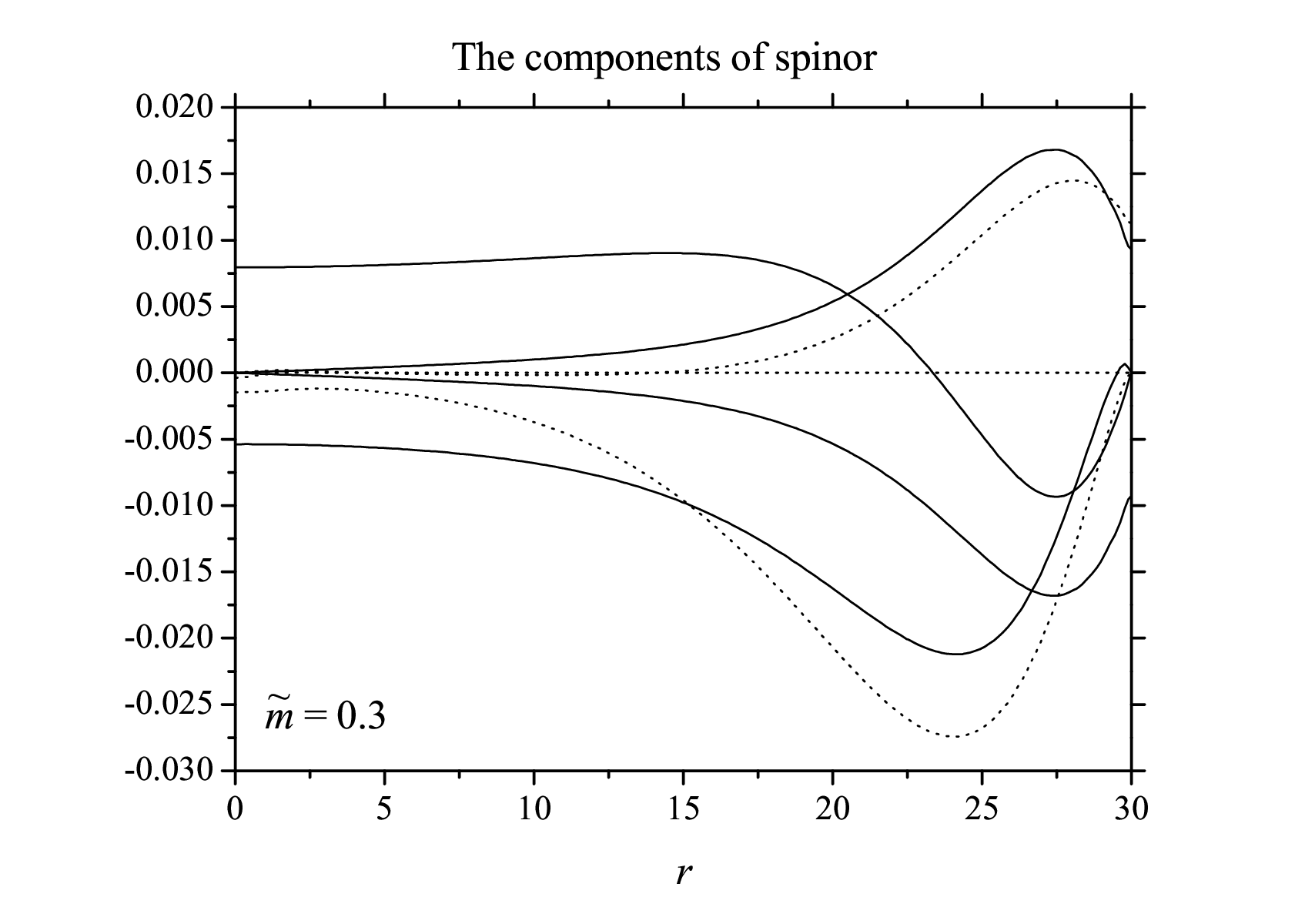}
\includegraphics[height=5.0cm, width=7.5cm]{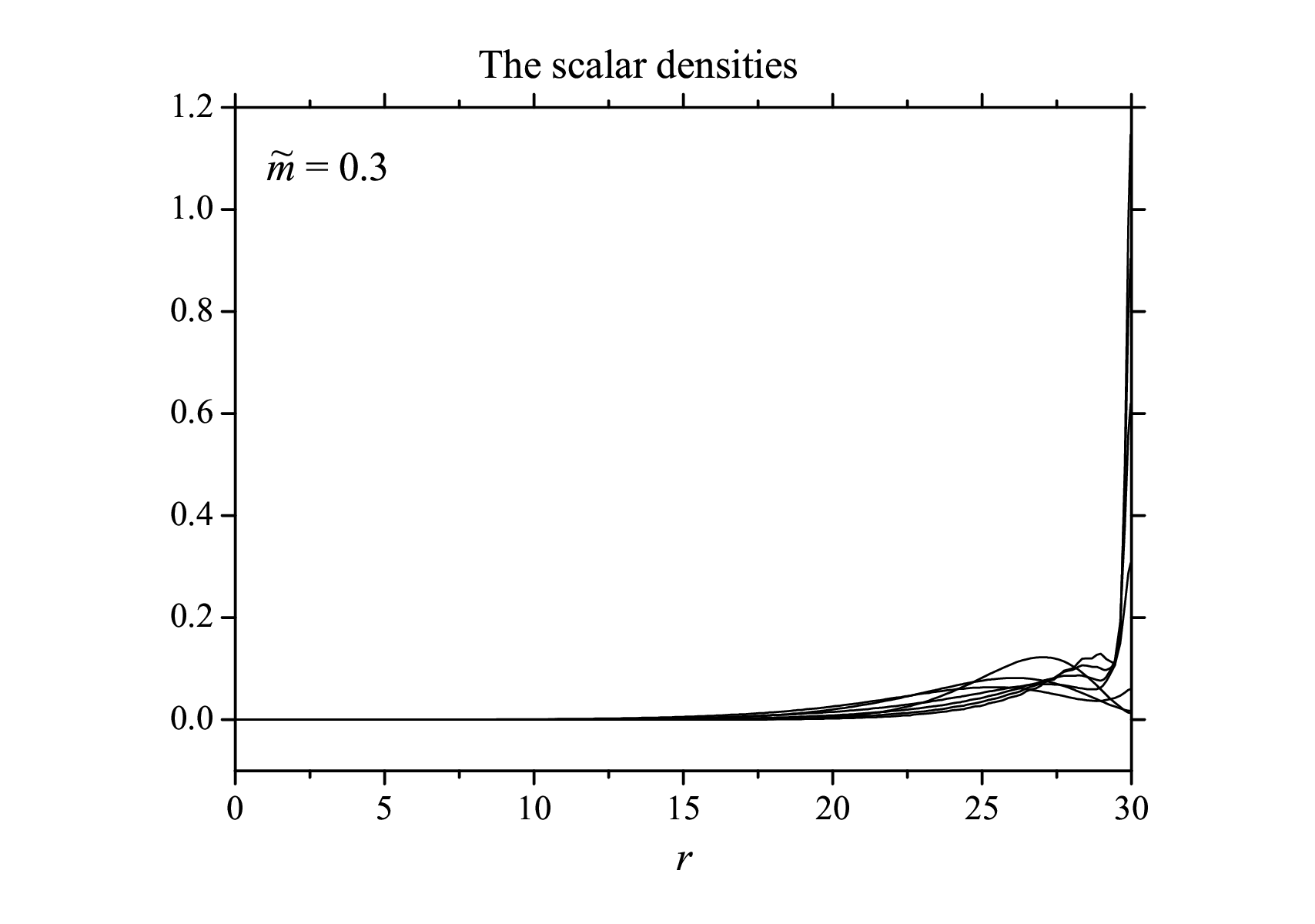}\\
\vspace{-1.05cm}
　\\
\includegraphics[height=5.0cm, width=7.5cm]{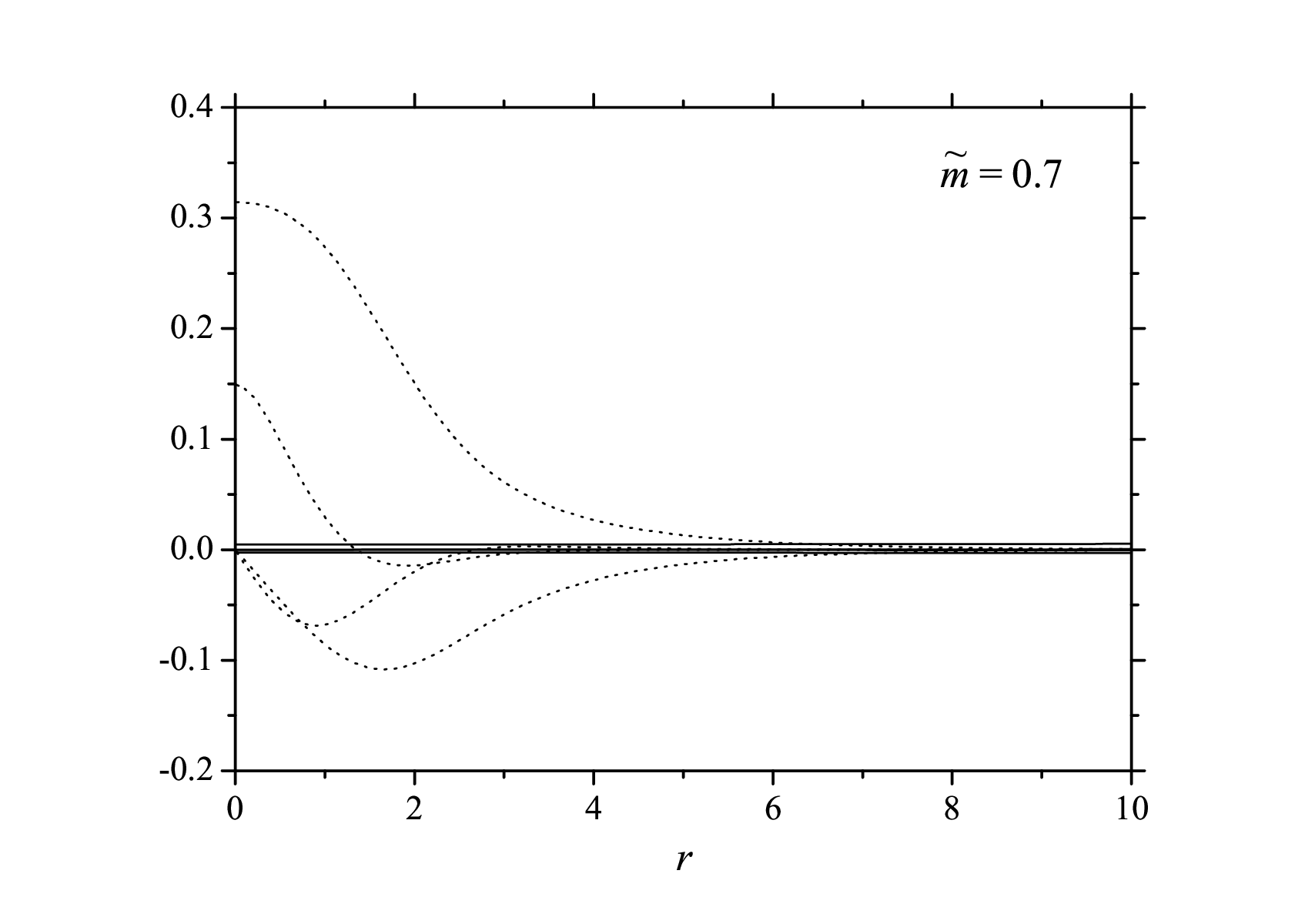}
\includegraphics[height=5.0cm, width=7.5cm]{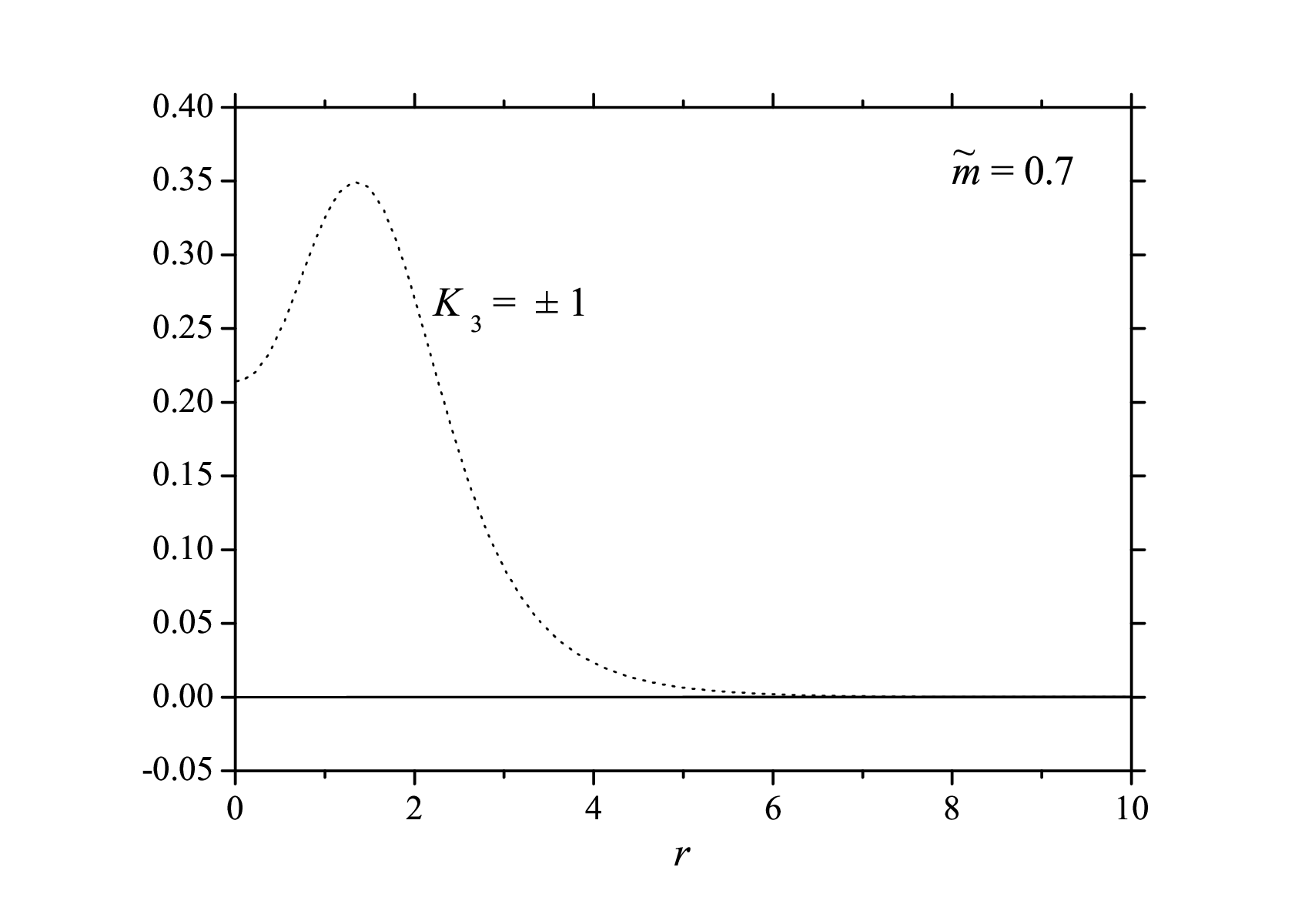}\\
\vspace{-1.05cm}
　\\
\includegraphics[height=5.0cm, width=7.5cm]{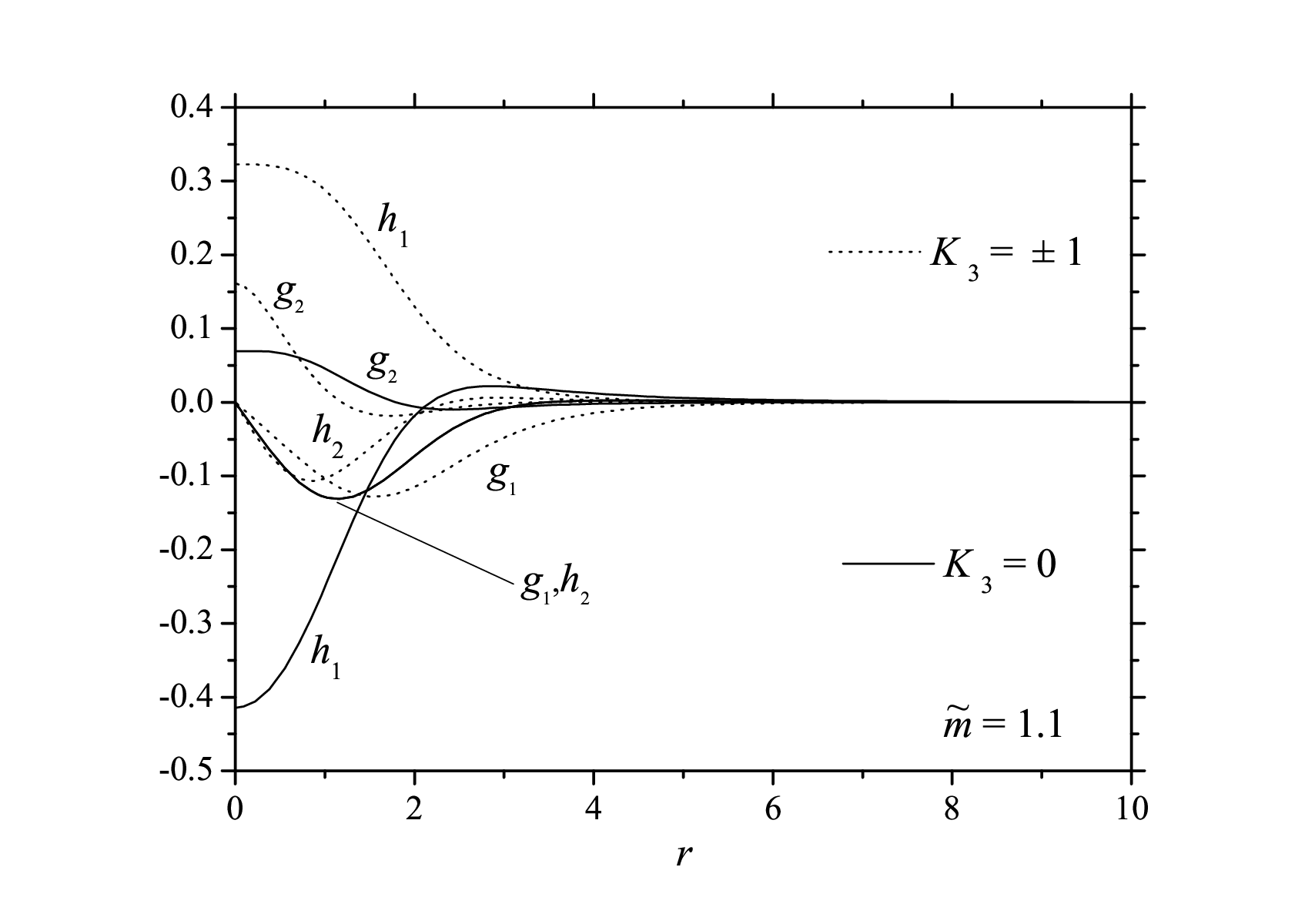}
\includegraphics[height=5.0cm, width=7.5cm]{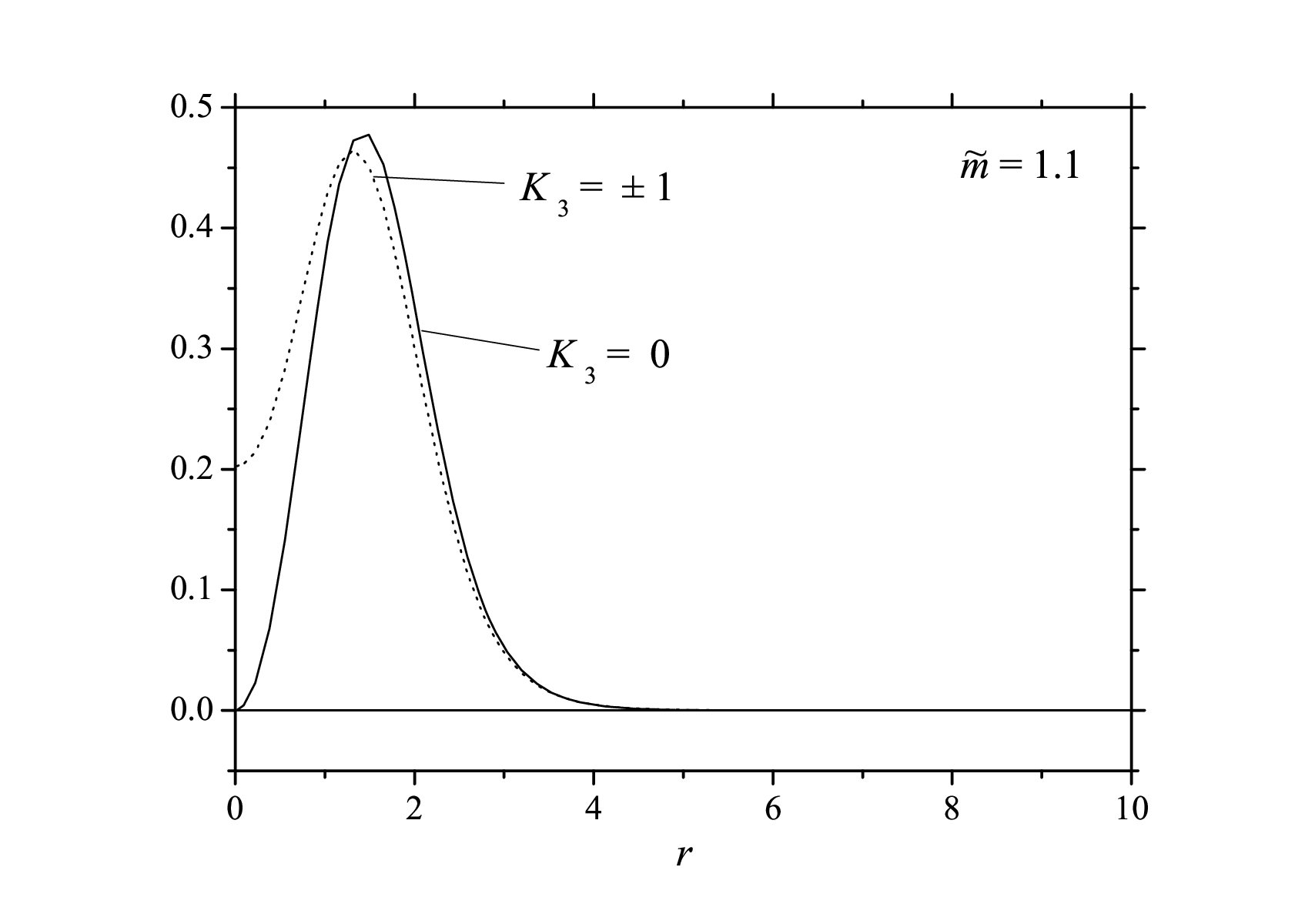}\\
\vspace{-1.05cm}
　\\
\includegraphics[height=5.0cm, width=7.5cm]{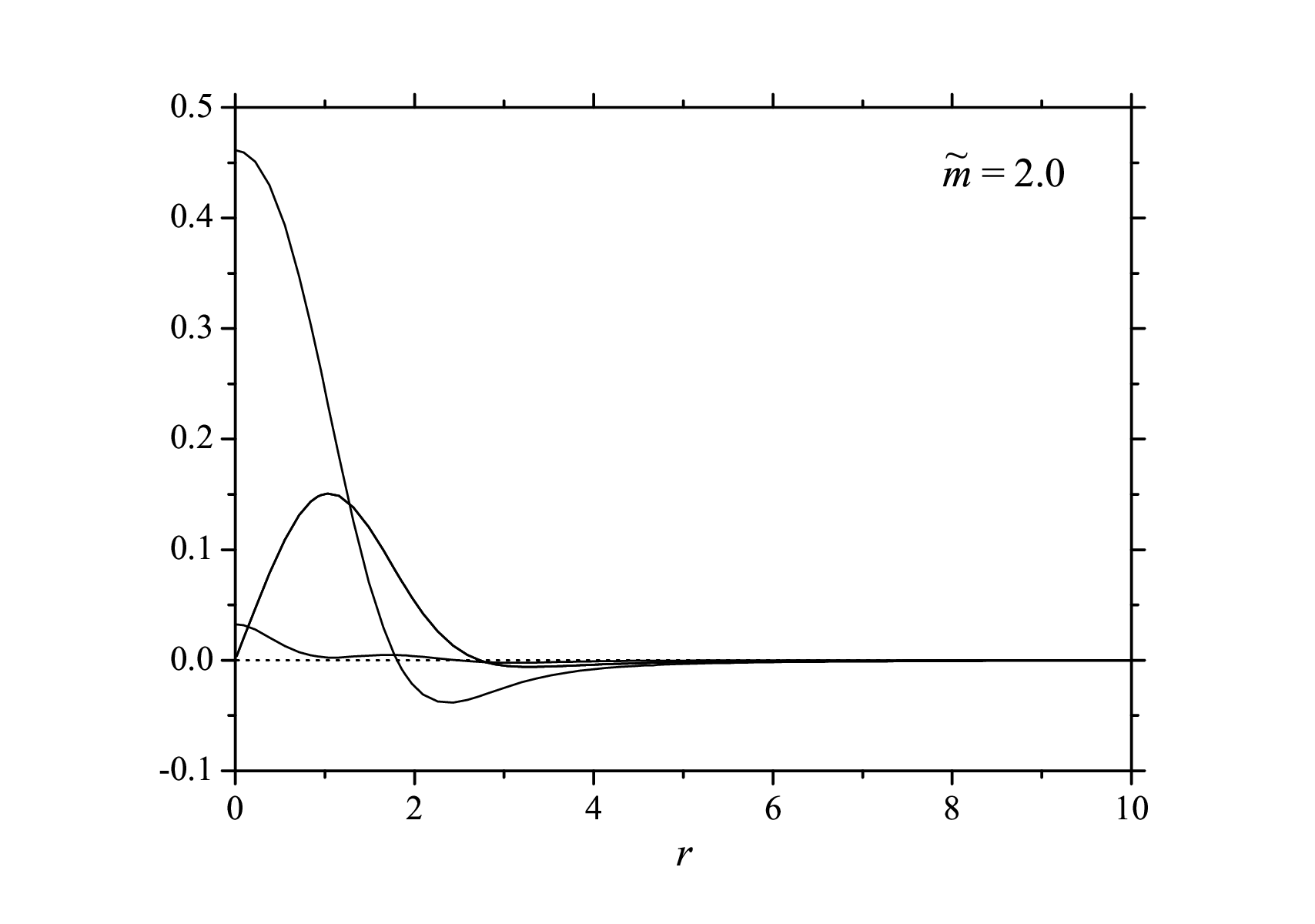}
\includegraphics[height=5.0cm, width=7.5cm]{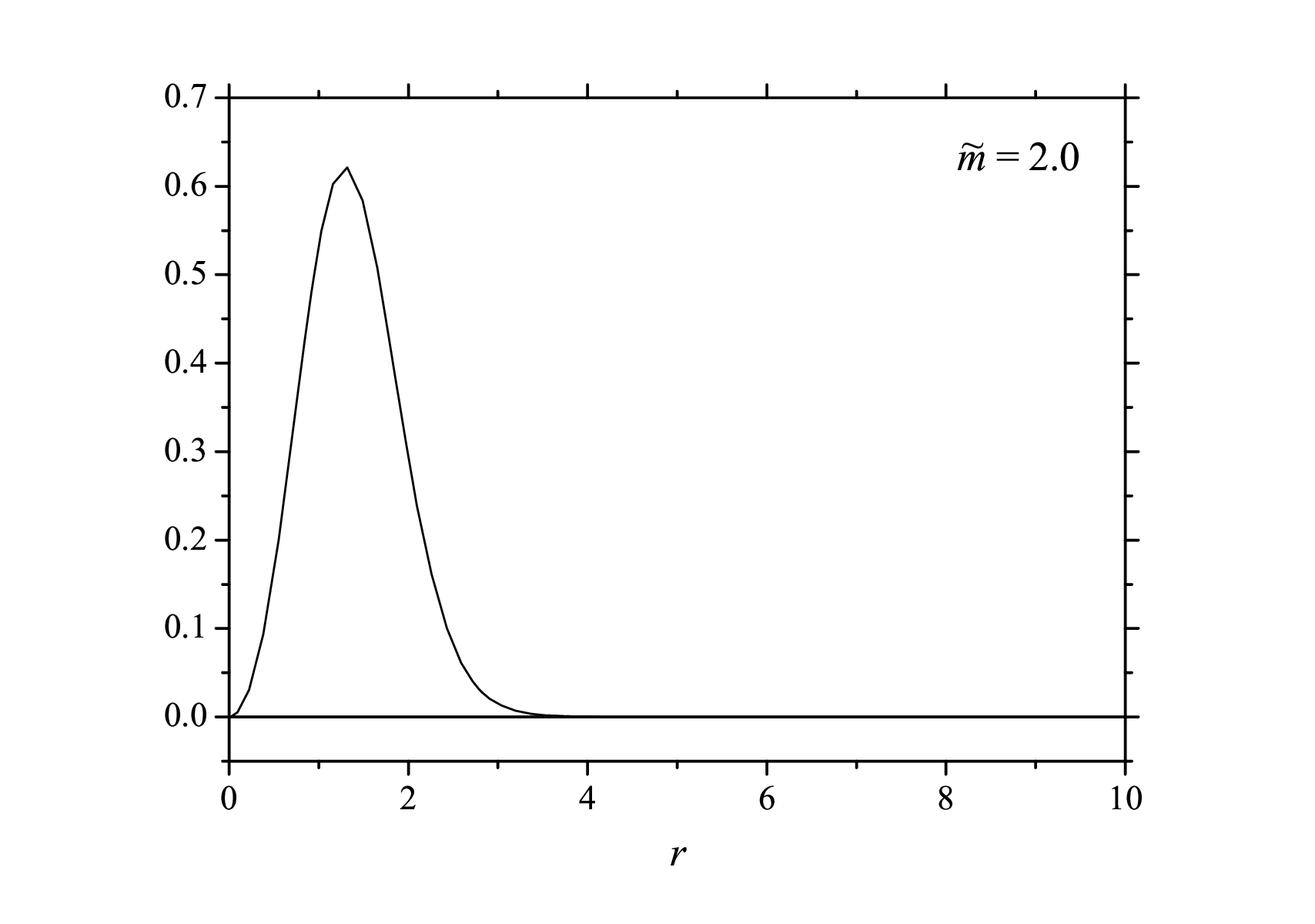}\\
\vspace{-1.05cm}
　\\
\includegraphics[height=5.0cm, width=7.5cm]{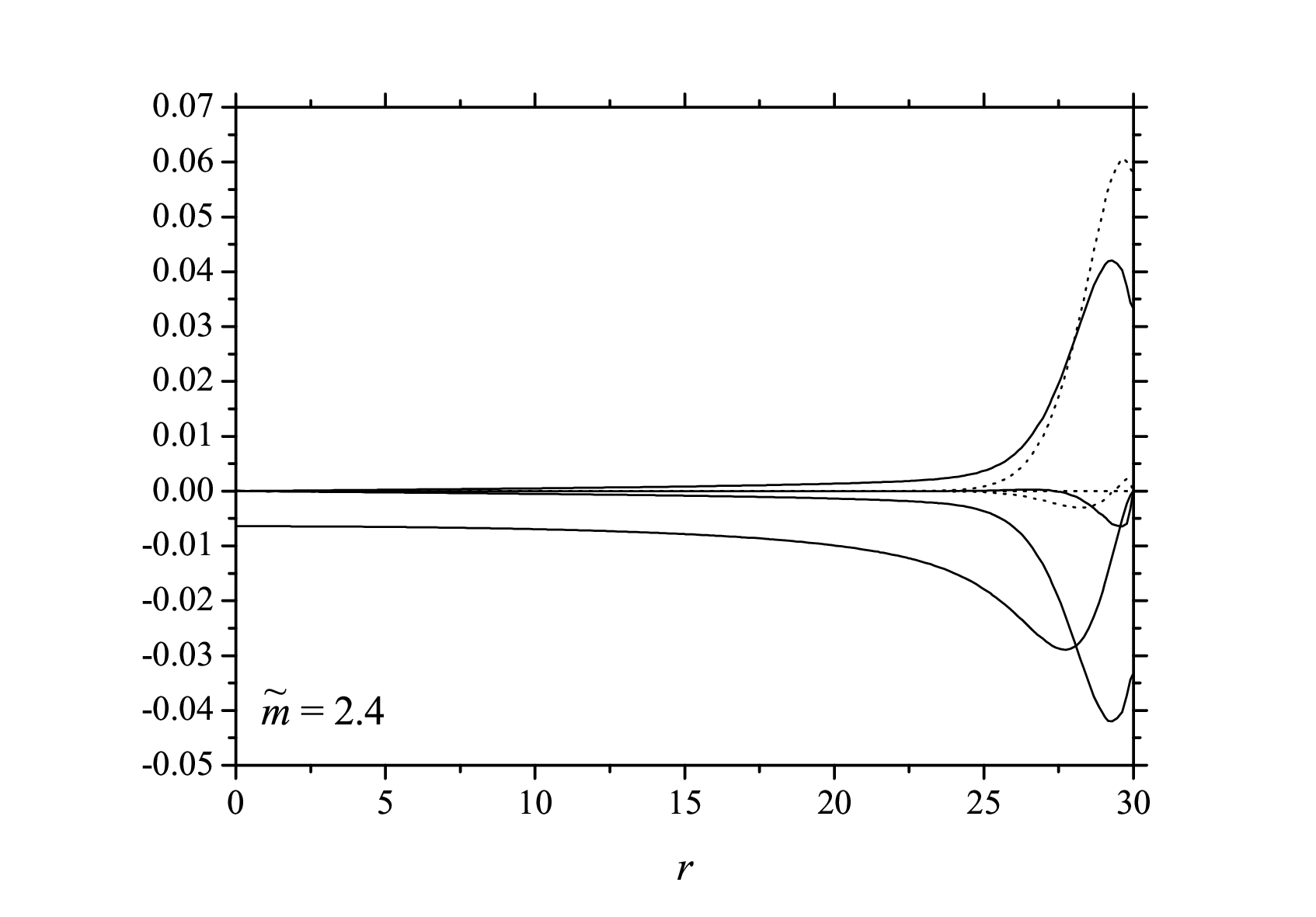}
\includegraphics[height=5.0cm, width=7.5cm]{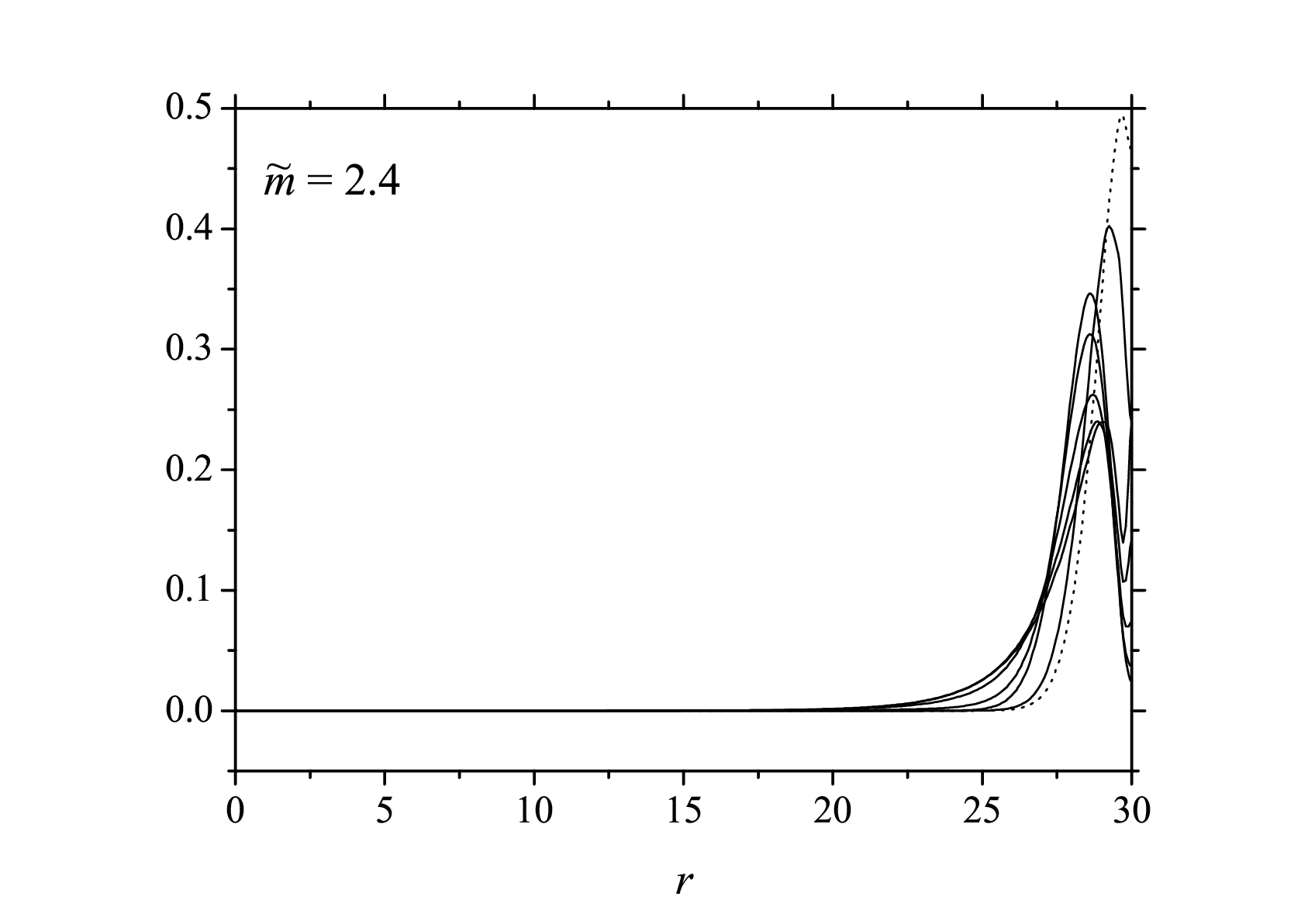}\\

\caption{\label{fig:wfsn3}The spinors (left) and the scalar densities (right) for the 
skyrmion backgrounds with $(m,n)=(1,3)$, $\alpha=0.01,\beta=-0.01,\gamma=0.005$
and $\mu=0.1$ for the several coupling constant $\tilde{m}$. A few of the components of the spinor 
sometimes are degenerate. }
\end{figure*}

In topological charge $N=3$, the index theorem states that three positive energy levels should dive to negative continuum, 
as shown in Fig. \ref{fig:spectralflow} for $(m,n)=(1,3)$ as well as for $(m,n)=(3,3)$.
However, in the latter case, the picture is much more sophisticated than for $m=1$. 
A doubly degenerated heavy state interchanges with a light single state at {\it junction A}.
The {\it junction B} is more complicated. At least, two isolated levels and two doubly degenerated levels
interact each other. 
(Similar behavior has already been observed in a somewhat different context in~\cite{Burnier:2006za}.)
After the spectral flow, one easily confirms that three levels dive from positive continuum to the negative.  
In the following analysis, we concentrate on the simpler case $(m,n)=(1,3)$ because it is more easy to get physical intuition.
Fig.\ref{fig:wfsn3} shows the localization properties of components of the spinors and the scalar 
density which is defined as 
\begin{eqnarray}
b_0(r)=\sum_{i=1}^2\int d\theta u_i^\dagger (r,\theta) u_i(r,\theta)
\end{eqnarray} 
for several coupling constant corresponding to $(m,n)=(1,3)$. 
We could produce the infinite tower of the spectra corresponding to $K_3=0,\pm 1,\pm 2 \cdots$ 
as well as the complete set of the wave vector.
Many of them form continuum (conducting level) and the lowest a few states are occupied levels. 
For small coupling constant, no localized solution exists. 
Increasing the coupling constant, the first degenerate states begin to localize and a single state follows. 
However, for a sufficiently large coupling constant, the spectra merges to the negative continuum and 
localized modes disappear. 

\begin{figure*}
\includegraphics[height=10.0cm, width=12.5cm]{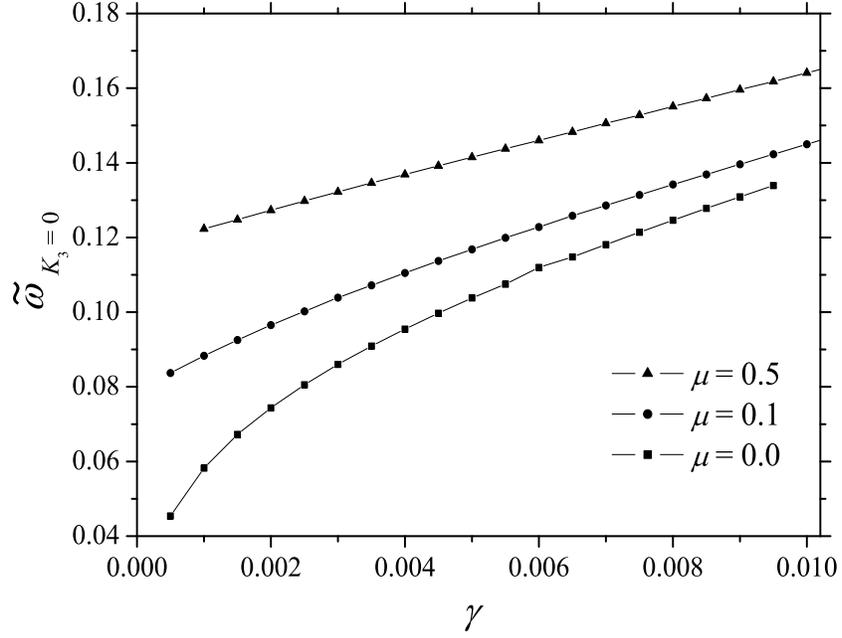}
\caption{\label{fig:massdif-gamma}The mass difference with $n=3$, $\alpha=0.05,\beta=0$.
}
\end{figure*}

\begin{figure*}
\includegraphics[height=10.0cm, width=12.5cm]{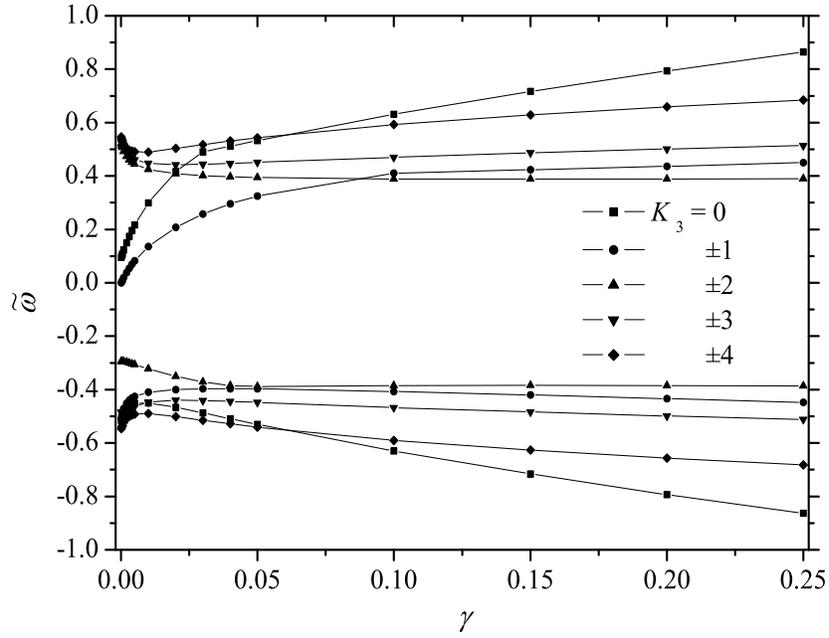}
\caption{\label{fig:mass-gamma}The spectral flow with $n=3$, $\alpha=0.0,\beta=0.0,\mu=0.1$, 
as a function of the cosmological constant.}
\end{figure*}

\begin{figure*}
\includegraphics[height=7.0cm, width=8.0cm]{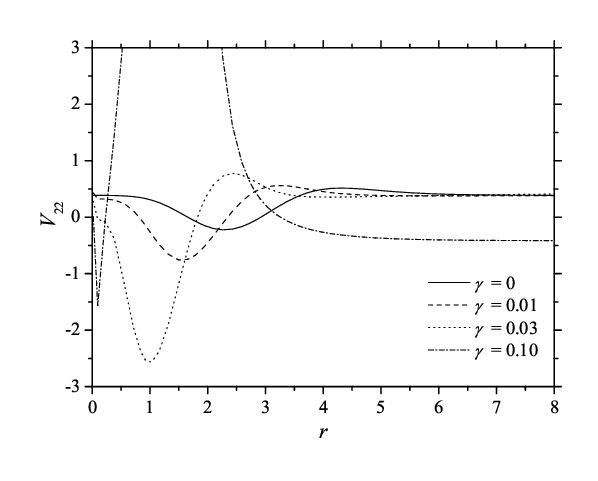}
\includegraphics[height=7.0cm, width=8.0cm]{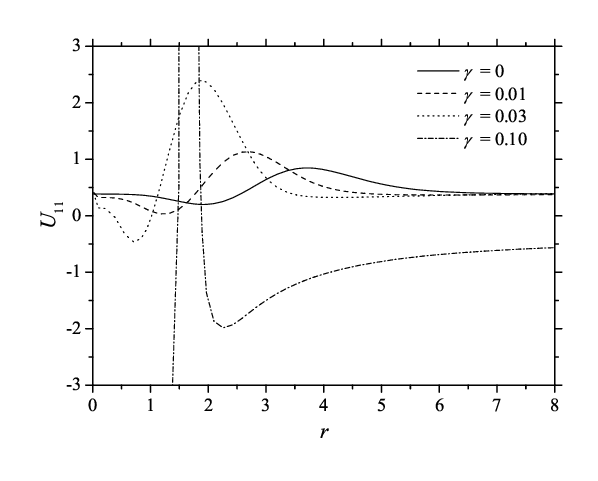}
\caption{\label{fig:potential-gamma}The effective potential $V_{22},U_{11}$ for several parameters $\gamma$ with 
the case of $n=3,\alpha=0.0,\beta=0.0,\mu=0.1$.
}
\end{figure*}

\section{Hubble parameter and the mass difference of the quarks}

The value of the Hubble parameter at the present epoch is about \cite{Riess:2011yx,Beutler:2011hx}
\begin{eqnarray}
H_0\sim 70~{\rm km~s^{-1} Mpc^{-1}}=1.5\times 10^{-39}~{\rm MeV}
\end{eqnarray}
In our previous analysis~\cite{Kodama:2008xm}, we used Skyrme parameters $\kappa_2,\kappa_4$ of order $\sqrt{\kappa_2/\kappa_4}\sim 10^{4}$ MeV, so the dimensionless parameter $\gamma_0$ is around
\begin{eqnarray}
\gamma_0=\frac{\kappa_4}{\kappa_2}H_0^2=10^{-8}\times (1.5\times 10^{-39})^2=2.3\times 10^{-86},
\end{eqnarray}
which is apparently negligible. 
However, during inflation the Hubble parameter should be quite different, i.e., 
the typical value is around $H_{\rm inf}\sim 10^{10}$GeV, which corresponds to $\gamma_{\rm inf}=10^{18}$ \cite{Cho:2004pc}.

In this section, we discuss the behavior of the massive fermion levels (quarks) for changing $\gamma$. 
Since we have several model parameters, we need to fix some of them in an appropriate way. 
We always get the localized modes for first/second degenerate levels (which corresponds to states $K_3=\pm 1$) by suitably adjusting the 
coupling constant $\tilde{m}$ for each $\gamma$.
If we simply set zero for these levels, the third generation (the $K_3=0$ state) also becomes localized mode. 
So we compute the mass of the third generation as the level of this localized mode $\tilde{\omega}_{K_3=0}$. 
In Fig.\ref{fig:massdif-gamma} we show the result for the case of $\mu=0.0,~0.1,~0.5$. The mass differences 
monotonically increase as Hubble parameter $\gamma$ grows. 
Note that the solutions with $\mu=0$ is naturally interpreted as corresponding to the case of $\kappa_0=0$, which is excluded at least in flat space by Derrick theorem. 
Here, we find solutions for $\beta=0$, it follows that in this case $\mu=0$ is to be regarded as other limits, 
compaptible with Derrick theorem. Such a limit can be achieved for $\kappa_2\to \infty~{\rm or}~\kappa_4=0$.

Roughly speaking, smaller values of $\gamma$ corresponds to late stages in the evolution of our Universe, in which case the mass 
difference monotonically decreases. In this sense, the parameter $\gamma$ allows to follow the properties of the fermions at 
different epochs during the inflation. 
However, our previous analysis \cite{Brihaye:2010nf} showed that there exists a maximal value of $\gamma$ where we could find solutions.
Typically the maximal values was of order $1$ which is extremely far from $\gamma_{\rm inf}$. It follows that there are two ways of interpreting our model:\\
~~(i)~our brane solution is suitable for describing only of the late stage in the evolution of our Universe;\\
~~(ii)~our brane solution could describes all age of Universe, but different values of the parameters
 $\kappa_2,\kappa_4$ should be employed for each epoch.
                                                                                                                                                                                                                                                                             
The first possibility (i) is the most natural interpretation. The value of $\gamma_{\rm crit}$ 
indicate that the solution starts with a late stage, 
which somehow makes sense since the model should be an effective model of some more fundamental mechanism.
Therefore the baby-skyrmion can only describes the quite recent Universe, already reducing speed of the inflation. 
Gap of the fermions between first/second and third generations gradually decreases as time grows (Fig.\ref{fig:massdif-gamma}). 
The change is, however, quite subtle. It depends on $\mu$, but the difference at the beginning is at most ten times larger than now. 

The second story (ii) seems strange. It is based upon a key assumption:
the model parameters of the baby-skyrmion are function of the age of the Universe.
We suppose that at early (or the beginning) stages, the value of $\gamma_{\rm inf}\equiv \gamma_{\rm crit}=2.8$. Therefore 
we estimate the model parameters at this epoch as
\begin{eqnarray}
\gamma_{\rm inf}=2.8=\frac{\kappa_4}{\kappa_2}\times 10^{26},~~
\therefore~~
\frac{\kappa_4}{\kappa_2}=2.8\times 10^{-26}.
\end{eqnarray}

In order to understand the structure of the brane in early time, it may be helpful to 
use known asymptotic solution of the baby-skyrmions in flat space of the form \cite{Piette:1994ug}:
\begin{eqnarray}
f(r)\sim\sqrt{\frac{\pi}{2}}\Bigl(\frac{\kappa_2}{\kappa_0}\Bigr)^{1/2}r^{-1/2}e^{-\sqrt{\kappa_0/\kappa_2}}~r
\end{eqnarray}
>From (\ref{eq:dimensionless2}) and one easily see
\begin{eqnarray}
\frac{\kappa_0}{\kappa_2}=\frac{\kappa_2}{\kappa_4}\mu = 3.6\times 10^{25}\mu
\end{eqnarray}
thus $f\to 0$ for this limit. This clearly indicates that the brane shrinks for the case of finite $\mu$, i.e. at early stages of inflation.
The dimensionful mass difference $\omega$ becomes
\begin{eqnarray}
\omega \equiv \sqrt{\frac{\kappa_2}{\kappa_4}}\tilde{\omega}\to 3.6\times 10^{25}\tilde{\omega}
\end{eqnarray}
which indicates that the third generation blows up. It rapidly decreases as time grows ($\kappa_2/\kappa_4$ decreases) and finally arrive 
to a finite value at our epoch ($\sqrt{\kappa_2/\kappa_4}\sim 10^4$). 

In conclusion, the mass difference decreases as time marches on, but in the picture (i), the effect is moderate so that the third generation
would be observable even at the early stage while in the picture (ii) it would never be observed because the energy is almost 
over $\sim 10^{20}$ than scope of any experimental facility. 
More detailed analysis is however, outside the scope of our present model. 

Next we investigate the broader range of the parameter $\gamma$. If we choose the coupling constant $\tilde{m}=0.62$ for a solution of
$\alpha=0.0,\beta=0.0,\mu=0.1$, it makes the first/second levels to zero at $\gamma=0$.
We could observe different type of the spectral flow (Fig.\ref{fig:mass-gamma}).
For decreasing value of $\gamma$, the levels $K_3=0,\pm 1$ dive from positive continuum and become localized modes. 
In this parameter case, the level $K_3=\pm 1$ begin to localize to the origin at $\gamma\sim 0.09$ and when $\gamma\sim 0.03$, the 
level $K_3=0$ becomes the localized mode. In Fig.\ref{fig:potential-gamma}, we plot diagonal part of the effective potentials $V_{22},U_{11}$
(the explicit forms are shown in the appendix, (\ref{epotential_v22}),(\ref{epotential_u11}).) Those exhibit the volcano form. For 
larger value of $\gamma$, the well becomes deep in depth but narrow in size. At a critical point (e.g.,$\gamma\sim 0.09$), 
the potential suddenly blows up and after that, no any localization modes appear. 

This mechanism explains creation of the particles (the massive fermions) in an expanding Universe. Generally, 
the particles appear by the pair creation/annihilation process. In the level crossing picture, the particle pair creation
occur during the level crossing the zero from negative to positive. 
In our case, however, the expanding brane captures and localizes the fermions
in the extra dimensional space time, which means that during inflation massive fermions suddenly appear on the brane 
(from upper continuum) and become lighter as $\gamma$ reduce, i.e. as inflation continues.

\section{Maxwell field on and off the brane}
In this section, we consider problem of a vector field in the background of the brane. 
According to the analysis for the 2+1 baby-Skyrme model with a U(1) gauge 
field \cite{gauge}, we introduce the action with the gauge field in six dimensions 
\be
S_{\rm Maxwell} = \int d^{6}x\sqrt{-g}\Bigl[ \frac{1}{4e} \mathcal F_{MN}\mathcal F^{MN} \Bigr]
\label{s_max}
\ee
and the usual derivative $\partial_M$ in the baby-Skyrme action $S_{\rm baby}$ is replaced to
the covariant derivative $\mathcal D_{M} = \partial_M  + A_M \bm n_\infty\times .~$ with the U(1) gauge field $A_M$. 
The $\mathcal F_{MN} := \partial_M A_N - \partial_N A_M$ is the Faraday tensor in six dimensions.
Variation of the action $S:=S_{\rm baby}+S_{\rm Maxwell}$ with respect to $A_M$ leads to
\bea
&& \nabla_M F^{MN} = e J^N,\nonumber\\
&& J^N = \left\{\kappa_2 (\bm n_\infty\times\bm n)\cdot (\mathcal D^{N}\bm{n}) - \kappa_4\left[(\bm n_\infty\times\bm n)\times\mathcal D_M\bm n\right].
\left[\mathcal D^N\bm n \times \mathcal D^M\bm n \right]\right\}.
\eea
We employ following parametrization for the gauge field
\be
A_M = (A_\mu(x^\nu) \tilde F(r,\theta) ,\ A_i (r,\theta)),
\label{ansa}
\ee
where $\mu,\nu$ refers to the four dimensional indices, i.e. $\mu,\nu=t,x,y,z$ and $i$ means the extradimensional
components such as $i,j=r,\theta$.
Using this parametrization, the Faraday tensor can be expressed as
\be
\mathcal  F_{AB} = \left(\begin{array}{c|c}
                \mathcal F^{(4)}_{\mu\nu} \tilde F 	& 	- A_\mu \partial_i \tilde F\\
 		\hline
		 A_\mu \partial_i \tilde F		&	\mathcal  F_{ij}
               \end{array}\right),
\ee
where $\mathcal F^{(4)}_{\mu\nu} = \partial_\mu A_\nu(x) - \partial_\nu A_\mu(x) $ is the four dimensional Faraday tensor.

Note that the action \eqref{s_max} contains a term $\frac{1}{4e}\tilde F^2 \mathcal F^{(4)}_{\mu\nu}\mathcal F^{(4)\mu\nu}$, 
which clearly means  
that not only the Planck mass, but also the electromagnetic coupling acquire the hierarchy. The effective four 
dimensional electromagnetic coupling would then become
\be
e_4 = \frac{e}{\int \tilde F^2(r,\theta) L(r)drd\theta}.
\ee
In fact, as we shall see in the following, the equation for $\tilde F$ turns out to be linear, thus 
we can choose the normalization of $\tilde F$ such that $\int \tilde F(r,\theta)^2 L(r)drd\theta=1$.

The equations for the gauge fields in the background of the baby skyrmion then become
\bea
\tilde F\nabla^\mu \mathcal F^{(4)}_{\mu\nu} + A_\nu(x)\Box \tilde F = eJ_\nu,  \\
-(\nabla^\mu A_\mu(x))\nabla_i \tilde F + \nabla^j \mathcal F_{ji}^{(2)} = e J_i.
\eea
The four dimensional components of the source $J_\mu$ 
is actually proportional to $A_\mu$ say $J_\mu=\tilde{J}(r,\theta) A_\mu$ where $\tilde{J}$ 
can be estimated via the extradimensional components, i.e. $f,f',M$ and $\tilde{F}$.
The equations for the extradimensional components then reduce to
\bea
&&\Box \tilde F-e \tilde{J}=\mu^2 \tilde F,\\
&&\nabla^jF_{ji}-eJ_i=\lambda\partial_i\tilde F,
\eea
where we used
\bea
&&\nabla^\mu \mathcal F^{(4)}_{\mu\nu} = -\mu^2 A_\nu,
\label{Mxwl} \\
&&\nabla^\mu A_\mu(x) = \lambda,
\label{Lgauge}
\eea 
where $\lambda, \mu$ are the variable separation constants. 
Quite interestingly, as a consequence of the variable separation \eqref{ansa}, 
above equations naturally implement the Maxwell equations and a gauge condition. 
>From a four dimensional point of view, the vector 
field should be massless, we impose $\mu=0$ and then (\ref{Mxwl}) is the four dimensional Maxwell equations. 
Also if we choose a special gauge choice $\lambda = 0$, (\ref{Lgauge}) is exactly the Lorentz gauge condition in four dimensions.
 
Note that direct computation shows that $J_\mu$ depends on the four dimensional
components only through a proportionality factor with $A_\mu$ and $J_i$ depends on the extradimensional coordinates only. Note also that the covariant 
derivative acting on four dimensional objects reduces to the four dimensional covariant derivative.

We can successfully decouple the four dimensional and extradimensional part when we impose the four dimensional Maxwell equations and 
the Lorentz gauge condition. 
We parameterize the extradimensional part as $A_i(r,\theta )dx^i = n A_\theta (r)d\theta $ and the extradimensional dependence of the four dimensional
components as $\tilde F(r,\theta) = F(r)$; which lead to the following two decoupled equations 
\bea
&&F'' + \left(2 \frac{M'}{M} + \frac{L'}{L}\right)F' +e F \left(f'^2+1\right) \sin ^2f=0,
\label{eqgaugee}\\
&&A_\theta'' +\left(4\frac{M'}{M}-\frac{L'}{L}\right)A'_\theta + e A_\theta \left(f'^2+1\right) \sin ^2f -e \left(f'^2+1\right) \sin ^2f=0.
\label{eqgauge4}
\eea
For the boundary conditions we impose
\be
F'(0)=0,\ A_\theta '(0)=0,F(0)=1, A_\theta(\infty) = 0,
\ee
where the first two are the regularity condition of the equations, the third one is an arbitrary choice of normalization and the last should be 
imposed in order to have no flux at 
infinity.

We should stress that all parameters do not lead to localizing gauge fields. The condition for localizing mode is of course the function $F$ decays 
to zero as $r$ grows. Since we impose that the four dimensional components are massless, the only parameter we can vary is actually the electromagnetic coupling.

For solving the equations, we use a standard fourth order Runge-Kutta method, with shooting for \eqref{eqgaugee} 
and the backward integration for \eqref{eqgauge4}. 
The reason for using backward integration is that it avoids the need of shooting to find the proper decay. 
As we shall discuss later, the source term in \eqref{eqgauge4} is non-trivial, and then it imposes the value of the function $A_\theta$ at the origin.

\subsection{Four dimensional gauge field confinement}
The near origin expansion of the four dimensional form function $F$ is given by
\be
F(r) = F_0 -\frac{F_0 f_n^2}{4 (n+1)^2} r^{2(n+1)} + \Ord{r}{2n+3},
\ee
where $F_0$ is a real constant and $f_n$ is the coefficient of the $r^n$ term in the near origin expansion of the function $f$.

We focus on the case of vanishing bulk cosmological constant. In this case, for large values of $r$ the function $F$ behaves like
\be
F(r)= F^{1}_\infty+\frac{F_\infty^2}{r} + \Ord{\frac{1}{r}}{2},
\ee
where $F_\infty^1,F_\infty^2$ are real constants that depend on the parameters, especially on $e$.

In principle, there should be a massive tower of four dimensional gauge fields. 
However, the lowest state is expected to be massless, so we concentrate on the zero mode. 
>From the fact is that not all values of the coupling lead to 
localized zero modes, i.e. modes for which the function
$F\rightarrow 0$ for large values of $r$ (in other words for which $F^1_\infty = 0$).

Our result indicates that only specific values of $e$ lead to localized four dimensional gauge fields. 
The process is somewhat similar to that of an eigenvalue 
problem, and in this case the coupling constant plays a role of the eigenvalue. In fact, when looking at \eqref{eqgaugee}, 
up to the sine terms, the equation certainly looks like an eigenvalue equation. 
The fact provides a clever quantization mechanism for the elementary electric charge due to the extradimensions.

\begin{figure}
 \includegraphics[scale=.4]{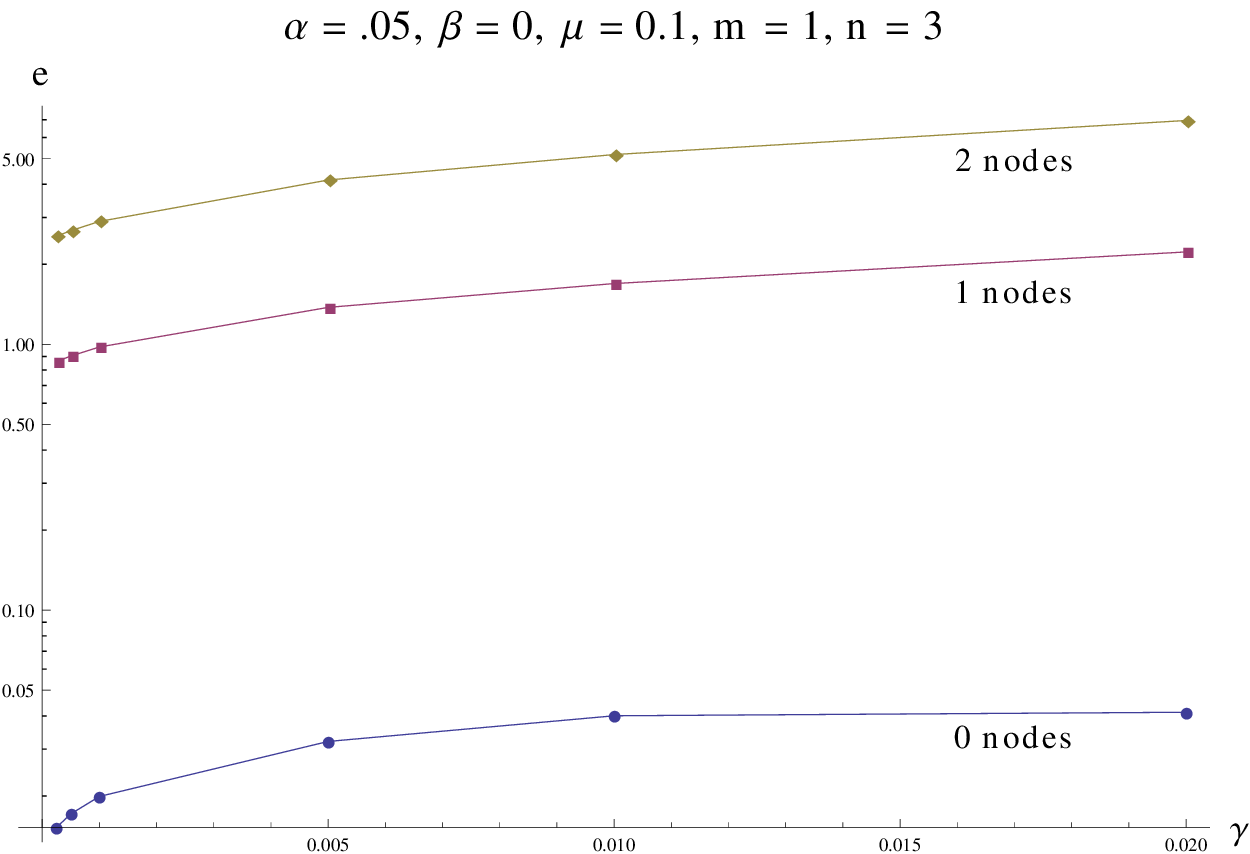}
\caption{The values of $e$ as a function of $\gamma$ for the first three localizing modes.}
\label{fig:e}
\end{figure}

\begin{figure}
 \includegraphics[scale=.3]{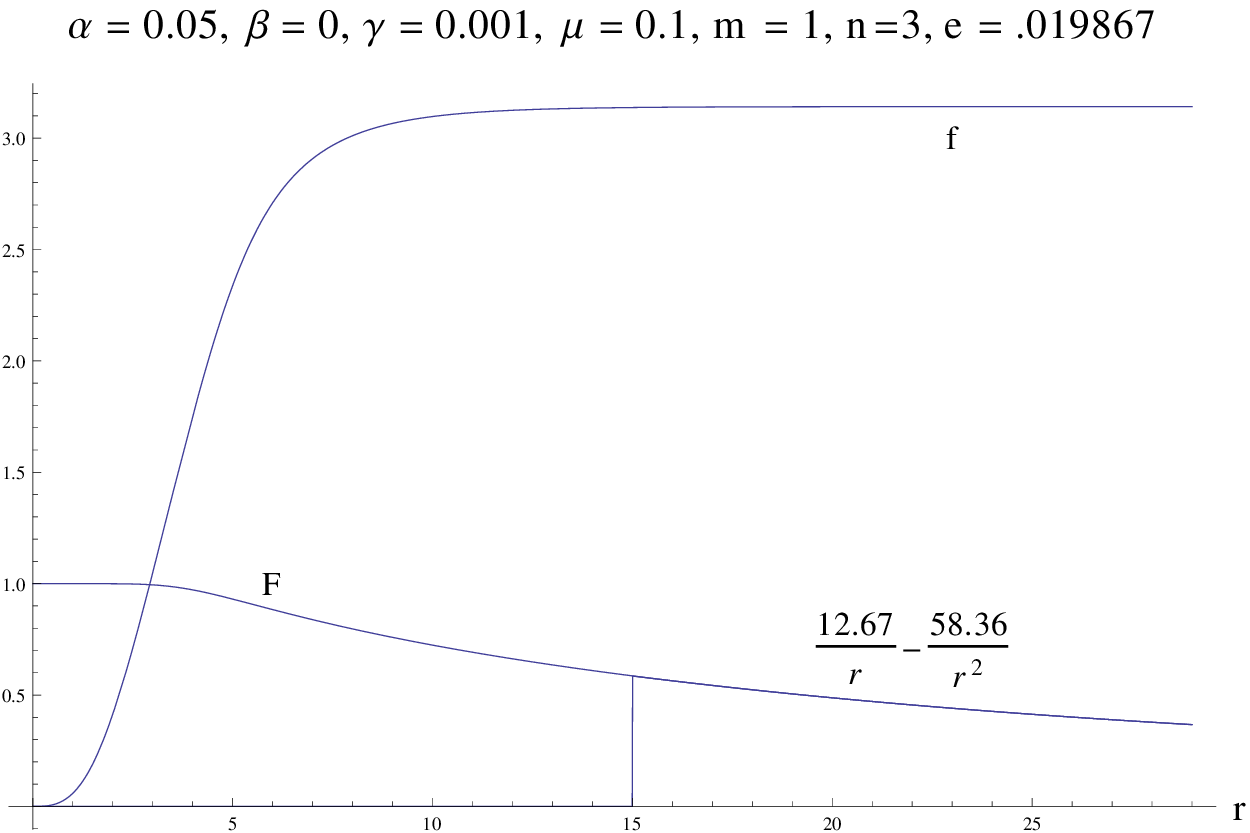}\hspace{.5cm}\includegraphics[scale=.3]{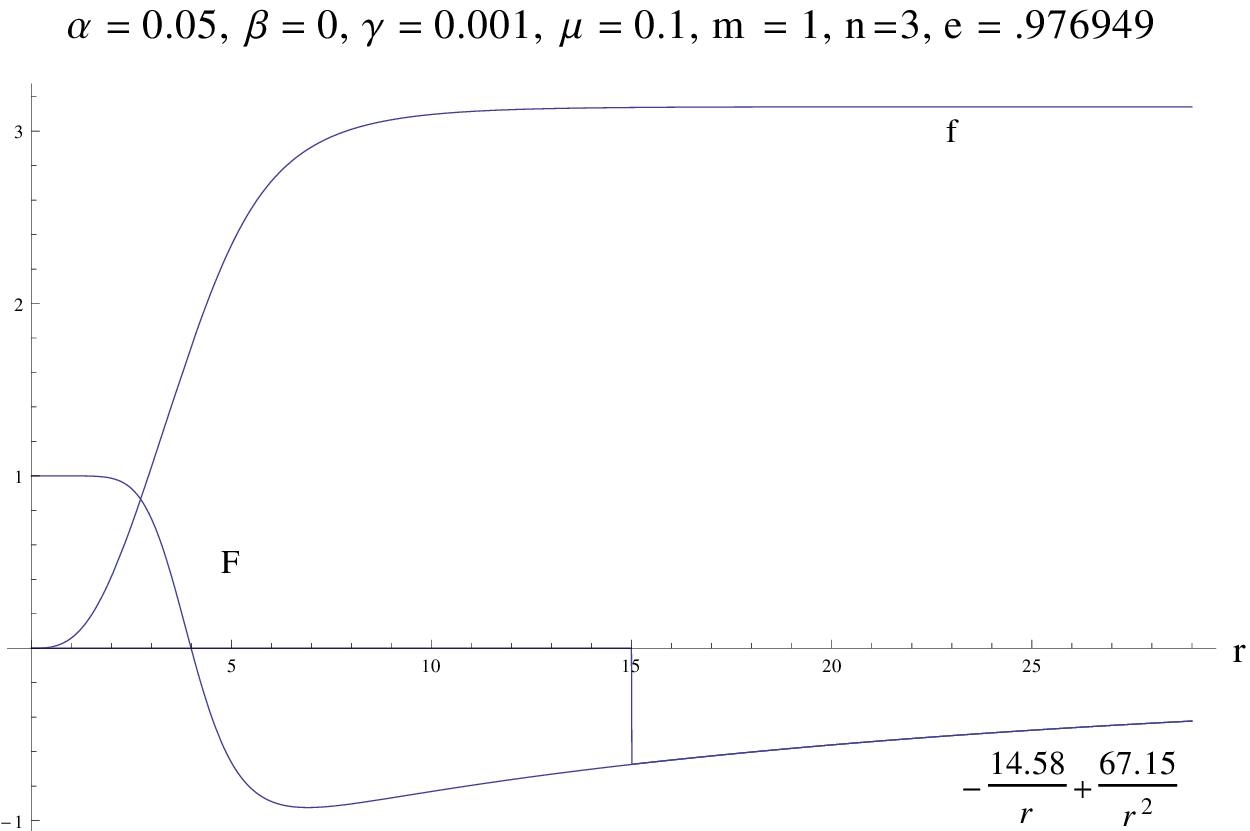}
\caption{The extradimensional dependance of the four dimensional components of the gauge field. We added the best fit of the decay, 
starting from the value $r=15$.}
\label{fig:form}
\end{figure}

Furthermore, we find that values of the electromagnetic coupling for the localized modes 
depend on the inflationary parameter. As discussed in the previous sections, varying the Hubble parameter in the model can be interpreted 
as looking to different time slices of the universe. Seen in this interpretation, our results suggest
that the value of the electric charge can evolve with time. This is illustrated on Fig. \ref{fig:e}. We also show two particular localized modes 
for specific values of the parameters in Fig. \ref{fig:form}.

\subsection{Gauge field in the extradimensions}
In this subsection, we consider the extradimensional component of the gauge field $A_\theta$ in \eqref{eqgauge4}.
The main difference of the equations is \eqref{eqgauge4} contains a source term. As a consequence, we cannot invoke the linearity to choose normalization,
instead, the source term imposes the value at the origin. One should keep in mind that we study the case of gauge field \emph{in the background} of the baby-Skyrme.
As a consequence, the results presented here might be affected by backreactions, especially in the case of larger electromagnetic couplings.

The near origin behaviour of the function $A_\theta$ is given by
\be
A_\theta =a_0+a_2 r^2+\frac{e f_n}{n (n+2)}r^{n+2}+\Ord{r}{n+4},
\ee
where $a_0, a_2$ are real constants and $f_n$ is again the coefficient of the $r^n$ term in the near origin expansion of the function $f$.
In the case of vanishing bulk cosmological constant and non-vanishing brane cosmological constant, the field $A_\theta$ decays according to 
\be
A_\theta = \frac{a_\infty}{r^3} + \Ord{\frac{1}{r}}{4}.
\ee

\begin{figure}
 \includegraphics[scale=.4]{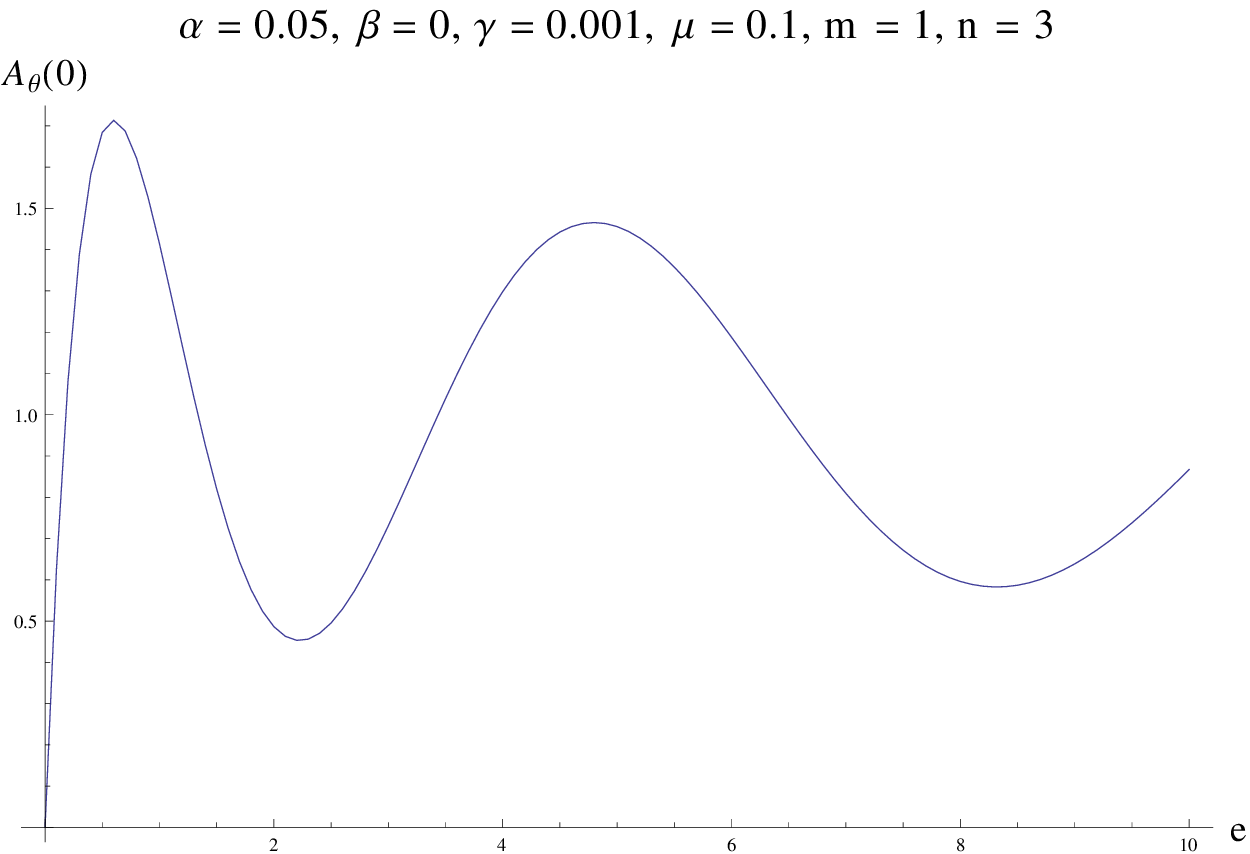}
\caption{The value at the origin of the extradimensional component of the gauge field as a function of the electromagnetic coupling.}
\label{fig:extra_A0}
\end{figure}

\begin{figure}
 \includegraphics[scale=.5]{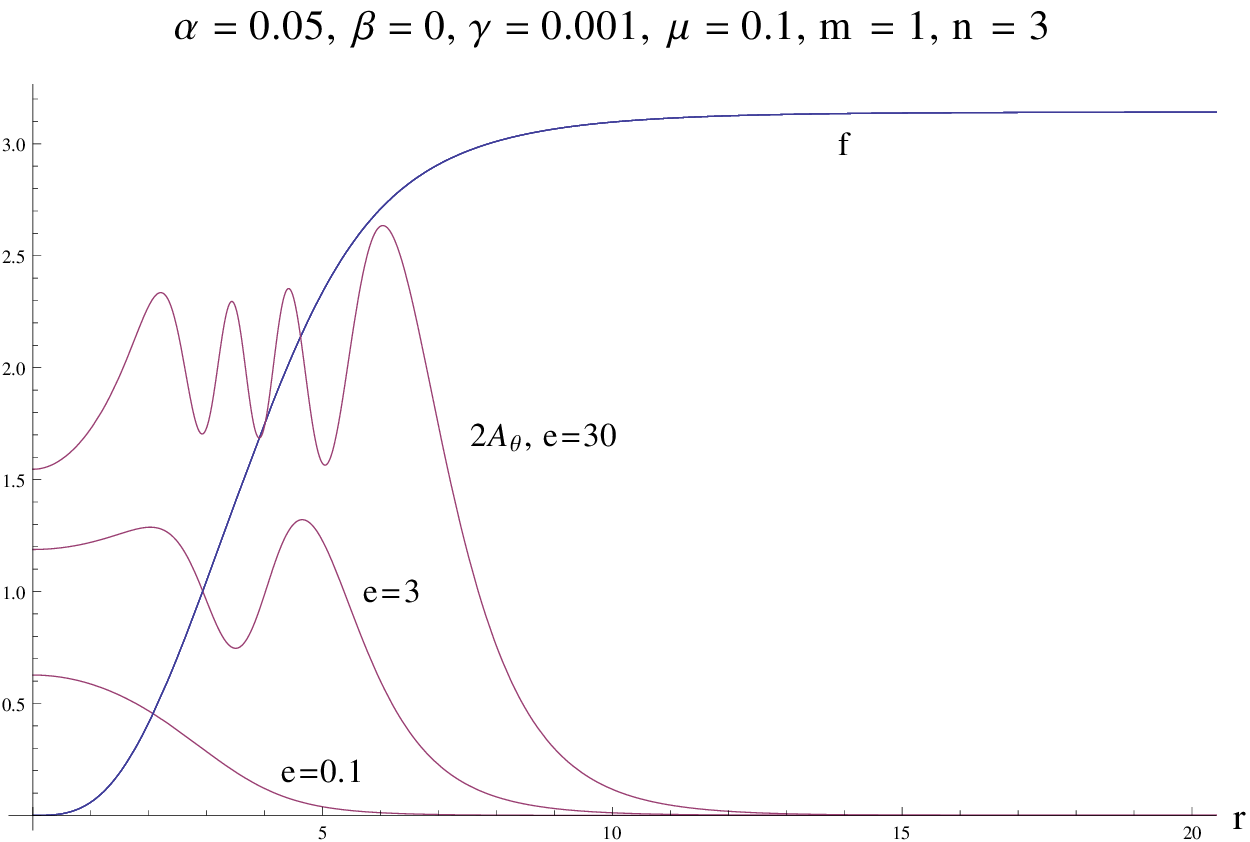}
\caption{The profile of the extradimensional component of the gauge field for three specific values of the coupling and specific values of the background
 parameter. The larger the coupling, the more the field is delocalized.}
\label{fig:extra_e}
\end{figure}

We find that for small values of the electromagnetic coupling, the extradimensional gauge field is more 
localized around the brane. As the coupling increases, the gauge field starts to produce a finite number of oscillations around some positive values before
decaying to zero. We note that the number of oscillations increases with the electromagnetic coupling. Furthermore, the more the field oscillates, the more
it is delocalized (i.e. the size grows). Note that the gauge field does not develop modes as it was doing in the four dimensional case; it 
always stays positive. We illustrate the above discussion on Figs. \ref{fig:extra_A0} and \ref{fig:extra_e}.

\newpage
\section{Summary}
We have studied fermion localization on the baby-skyrmion brane with positive cosmological constant. 
We especially have concentrated on the solutions with topological charge $N=3$. 
The matter distributions have a ring shaped but for $m=3$ it is more concentrate on a region close to the origin. 
The fermions coupled with the baby-Skyrme field has an isospin doublet structure, thus we naturally could take 
into account the flavor symmetry of the quarks. In terms of the index theorem, number of the fermions localized on
the brane coincides with the topological charge $N$. For $N=3$, we obtained the three localized modes during
a parameter range of the coupling constant $\tilde{m}$. The doubly degenerate levels are always lower than the 
single excited state, the levels are occupied by quarks/leptons of lower two generations.
Especially we studied about the mass difference between the first two generations and the third generation 
for changing dimensionless Hubble parameter $\gamma$. The behavior exhibits typical spectral flow. For decreasing 
value of $\gamma$, the three localized modes appear from the upper continuum when $\gamma\sim 0.03$,  
which indicates new particle creation mechanism. 

Our scheme is symmetric to the flavor (isospin) degrees of freedom thus $(u,c,t)$ sector and $(d,s,b)$ sector are degenerate.
In order to split the degeneracy, we introduce explicit symmetry breaking of the skyrmion 
parameters $\kappa_2,\kappa_4$. For $(d,s,b)$ sector the choice $\sqrt{\kappa_2/\kappa_4}\sim 10^4$ MeV 
reproduce the mass difference between the first/second and the third generationsm while 
for $(u,c,t)$ sector we adopt $\sqrt{\kappa_2/\kappa_4}\sim 10^6$ MeV to achieve an agreement with observation. 
This prescription equals to introduce the flavor hierarchical structure to the coupling constant $m$ (for a fixed $\tilde{m}$)
in terms of the relation  (\ref{dimensionless_mw}).  
On the other hand, in Ref.\cite{Kodama:2008xm} 
one of the authors of the present paper considered the breaking of the time-reversal symmetry by 
taking into account the rotational symmetry breaking of the baby-skyrmions. The authors of \cite{Kodama:2008xm} 
introduced an arbitrary deformation parameter by hand which successfully reproduce the mass splitting of the first/second generations. 
Recently we have found the existence of the $\mathbb{Z}_2$ symmetric, deformed brane in the baby-Skyrme model
\cite{Delsate:2012}. It is worth to computing the spectra of the fermions coupled to the deformed brane. 
As a conclusion, for the flavor hierarchy we could solve by introducing the hierarchical coupling constant of the 
fermions and the baby-skyrmions. Also the geometrical structure of the brane (background baby-skyrmions) brings 
about a solution to the mass hierarchy within the family. A more quantitative analysis implementing all these issues 
will be discussed in the forthcoming paper. 

In this paper, we employed left/right symmetric coupling scheme of the fermions and the baby-skyrmions. And we 
used the six dimensional generalization of the Dirac-Pauli representation for the gamma matrices. Although
there are several advantages to computing the massive modes, the mechanism of localization of chiral fermions 
on the brane is still absent. The study is mandatory to achieve full understanding on properties of our SM particles
and will be reported in near future. 
 
On another hand, we studied the localization properties of gauge fields and found that indeed, it is possible to localize four dimensinal gauge fields
on the brane for very specific values of the electromagnetic coupling. We studied the gauge field in the background of the brane, i.e. without 
backreactions. Although we believe that it is already quite instructive for the four dimensional part of the vector field, we think that
the extradimensional components of the gauge fields should be considered with full backreactions. The good news is that, at least in our approach, the 
extradimensional sector and the four dimensional sector of the vector field decouple. 

We forced massless gauge fields, even if it is well known that higher dimensions usually have the effect of producing mass towers for the fields of interest.
In our case, we know that electromagnetic gauge fields are massless from a four dimensional point of view. This implies that we should at least have a massless
mode for the four dimensional gauge fields. This zero mode should actually be the first state of a massive higher energy tower of vector fields (which we do not
study in this paper). 

As an plausible extension of the present study is to consider non-abelian gauge group confinement on solitonic branes. As we observed in this paper, 
imposing existence of massless vector mode automatically introduces quantization on the coupling, 
which is somehow the inverse of the usual approach, namely, compute masses for a given coupling
in order to find localized modes. This is certainly valid for a low energy approximation of massless vectors, such as $U(1)$ fields.
However, for vector bosons like SU(2), provided the $W,Z$ particles, it does not work because they have observed masses. 
One should then either recall the standard procedure of computing the mass for given coupling or find a new mechanism that generates mass and predicts coupling.

\appendix
\section{The components for the effective potential}

The $V_{ij},W_i~(i,j=1,2)$ in (\ref{2diffeq-a}) are summarized as
\begin{eqnarray}
&&V_{11}=\frac{1}{2}A'_{11}+\frac{1}{4}\Bigl(A_{11}+\frac{1}{L}\Bigr)^2+A_{11}\frac{l}{L}-\frac{\omega^2}{M^2}+\tilde{m}^2 \nonumber \\
&&\hspace{1cm}-2\frac{M''}{M}-2\frac{M'^2}{M^2}-2\frac{M'}{M}\frac{L'}{L}-\frac{1}{2}\frac{L''}{L}+\frac{1}{4}\frac{L'^2}{L^2}
-\frac{1}{2}\frac{L'}{L^2}+\frac{l(1-L')}{L^2}\,, \\
&&V_{12}=A_{12}\biggl(-\frac{1}{2}\Bigl(A_{22}+\frac{1}{L}\Bigr)+2\frac{M'}{M}+\frac{1}{2}\frac{L'}{L}-\frac{l+n}{L}\biggr)
e^{-\frac{1}{2}\int^r (A_{22}-A_{11})dr'}\,, \\
&&V_{21}=A_{21}\biggl(-\frac{1}{2}\Bigl(A_{11}+\frac{1}{L}\Bigr)+2\frac{M'}{M}+\frac{1}{2}\frac{L'}{L}-\frac{l}{L}\biggr)
e^{-\frac{1}{2}\int^r (A_{22}-A_{11})dr'}\,, \\
&&V_{22}=\frac{1}{2}A'_{22}+\frac{1}{4}\Bigl(A_{22}+\frac{1}{L}\Bigr)^2+A_{22}\frac{l+n}{L}-\frac{\omega^2}{M^2}+\tilde{m}^2 \nonumber \\
&&\hspace{1cm}-2\frac{M''}{M}-2\frac{M'^2}{M^2}-2\frac{M'}{M}\frac{L'}{L}-\frac{1}{2}\frac{L''}{L}+\frac{1}{4}\frac{L'^2}{L^2}
-\frac{1}{2}\frac{L'}{L^2}-\frac{(l+n)(1-L')}{L^2}\,,
\label{epotential_v22}
\\
&&W_1=A_{12}e^{-\frac{1}{2}\int^r (A_{22}-A_{11})dr'}\,, \\
&&W_2=A_{21}e^{-\frac{1}{2}\int^r (A_{11}-A_{22})dr'}\,.
\end{eqnarray}
For (\ref{2diffeq-b}), the components $U_{ij},(i,j=1,2)$ are written as
\begin{eqnarray}
&&U_{11}=\frac{1}{2}A'_{22}+\frac{1}{4}\Bigl(A_{22}+\frac{1}{L}\Bigr)^2+A_{22}\frac{l+1}{L}-\frac{\omega^2}{M^2}+\tilde{m}^2\,, \nonumber 
\label{epotential_u11}\\
&&\hspace{1cm}-2\frac{M''}{M}-2\frac{M'^2}{M^2}-2\frac{M'}{M}\frac{L'}{L}-\frac{1}{2}\frac{L'}{L}+\frac{1}{4}\frac{L'^2}{L^2}
-\frac{1}{2}\frac{L'}{L^2}+\frac{(l+1)(L'-1)}{L^2}\,,\\
&&U_{12}=\tilde{A}_{12}\biggl(-\frac{1}{2}\Bigl(A_{11}+\frac{1}{L}\Bigr)+2\frac{M'}{M}+\frac{1}{2}\frac{L'}{L}+\frac{l+n+1}{L}\biggr)
e^{-\frac{1}{2}\int^r (A_{11}-A_{22})dr'}\,, \\
&&U_{21}=\tilde{A}_{21}\biggl(-\frac{1}{2}\Bigl(A_{22}+\frac{1}{L}\Bigr)+2\frac{M'}{M}+\frac{1}{2}\frac{L'}{L}+\frac{l+1}{L}\biggr)
e^{-\frac{1}{2}\int^r (A_{22}-A_{11})dr'}\,, \\
&&U_{22}=\frac{1}{2}A'_{11}+\frac{1}{4}\Bigr(A_{11}+\frac{1}{L}\Bigr)^2+A_{11}\frac{l+n+1}{L}-\frac{\omega^2}{M^2}+\tilde{m}^2 \nonumber \\
&&\hspace{1cm}-2\frac{M''}{M}-2\frac{M'^2}{M^2}-2\frac{M'}{M}\frac{L'}{L}-\frac{1}{2}\frac{L''}{L}+\frac{1}{4}\frac{L'^2}{L^2}
-\frac{1}{2}\frac{L'}{L^2}+\frac{(l+n+1)(L'-1)}{L^2}\,.
\end{eqnarray}
The $A_{ij} (i,j=1,2)$ are obtained as
\begin{eqnarray}
&&A_{11}:=-\biggl(\frac{\omega^2}{M^2}-\tilde{m}^2\biggr)^{-1}
\biggl(\tilde{m}^2\frac{n}{L}\sin^2f-\frac{M'\omega^2}{M^3}-\frac{\omega \tilde{m}}{M}\sin ff'+\frac{M'\omega\tilde{m}}{M^2}\cos f\biggr)\,, \nonumber \\
&&A_{12}:=\Bigl(\frac{\omega^2}{M^2}-\tilde{m}^2\Bigr)^{-1}
\biggl(\frac{\tilde{m}\omega}{M}\sin f\Bigl(\frac{n}{L}-\frac{M'}{M}\Bigr)+\tilde{m}\cos f\Bigl(\frac{n\tilde{m}}{L}\sin f-\frac{\omega f'}{M}\Bigr)-\tilde{m}^2 f'\biggr)\,, \nonumber \\
&&A_{21}:=-\Bigl(\frac{\omega^2}{M^2}-\tilde{m}^2\Bigr)^{-1}
\biggl(\frac{\tilde{m}\omega}{M}\sin f\Bigl(\frac{n}{L}+\frac{M'}{M}\Bigr)-\tilde{m}\cos f\Bigl(\frac{n\tilde{m}}{L}\sin f-\frac{\omega f'}{M}\Bigr)-\tilde{m}^2 f'\biggr)\,,\\
&&A_{22}:=\biggl(\frac{\omega^2}{M^2}-\tilde{m}^2\biggr)^{-1}
\biggl(\tilde{m}^2\frac{n}{L}\sin^2f+\frac{M'\omega^2}{M^3}-\frac{\omega \tilde{m}}{M}\sin ff'+\frac{M'\omega\tilde{m}}{M^2}\cos f\biggr)\,, \nonumber 
\label{a-components}
\end{eqnarray}
and the $\tilde{A}_{12},\tilde{A}_{21}$ can be estimated similar to $A_{ij}$, and are found to be
\begin{eqnarray}
&&\tilde{A}_{12}:=\Bigl(\frac{\omega^2}{M^2}-\tilde{m}^2\Bigr)^{-1}
\biggl(\frac{\tilde{m}\omega}{M}\sin f\Bigl(\frac{n}{L}+\frac{M'}{M}\Bigr)-\tilde{m}\cos f\Bigl(\frac{n\tilde{m}}{L}\sin f-\frac{\omega f'}{M}\Bigr)-\tilde{m}^2 f'\biggr)\,, \\
&&\tilde{A}_{21}:=-\Bigl(\frac{\omega^2}{M^2}-\tilde{m}^2\Bigr)^{-1}
\biggl(\frac{\tilde{m}\omega}{M}\sin f\Bigl(\frac{n}{L}-\frac{M'}{M}\Bigr)+\tilde{m}\cos f\Bigl(\frac{n\tilde{m}}{L}\sin f-\frac{\omega f'}{M}\Bigr)-\tilde{m}^2 f'\biggr)\,.
\end{eqnarray}

\end{document}